\documentclass[journal=jacsat,manuscript=article]{achemso}
\usepackage[version=3]{mhchem} %
\usepackage{graphicx} %
\usepackage{dsfont}
\usepackage{amsmath}
\usepackage{amssymb}
\usepackage{xcolor}
\usepackage{multirow}
\usepackage{hyperref}
\usepackage{dsfont}
\usepackage[normalem]{ulem}
\usepackage{array}
\usepackage{booktabs} %
\newcommand{\COMMENTED}[1]{}

\newcommand{\CHANGES}[1]
{#1}

\SectionNumbersOn

\author{Siyuan Chen}
\affiliation{Pritzker School of Molecular Engineering, University of Chicago, Chicago, Illinois 60637, United States}
\author{Victor Wen-zhe Yu}
\affiliation{Materials Science Division, Argonne National Laboratory, Lemont, Illinois 60439, United States}
\author{Yu Jin}
\affiliation{Pritzker School of Molecular Engineering, University of Chicago, Chicago, Illinois 60637, United States}
\author{Marco Govoni}
\affiliation{Department of Physics, Computer Science, and Mathematics, University of Modena and Reggio Emilia, Modena, 41125, Italy}
\author{Giulia Galli}
\affiliation{Pritzker School of Molecular Engineering, University of Chicago, Chicago, Illinois 60637, United States}
\alsoaffiliation{Department of Chemistry, University of Chicago, Chicago, Illinois 60637, United States}
\alsoaffiliation{Materials Science Division, Argonne National Laboratory, Lemont, Illinois 60439, United States}
\email{gagalli@uchicago.edu}

\title{Advances in quantum defect embedding theory}

\abbreviations{IR,NMR,UV}
\begin{document}

\begin{tocentry}
\includegraphics[width=8.24cm,height=3.8625cm]{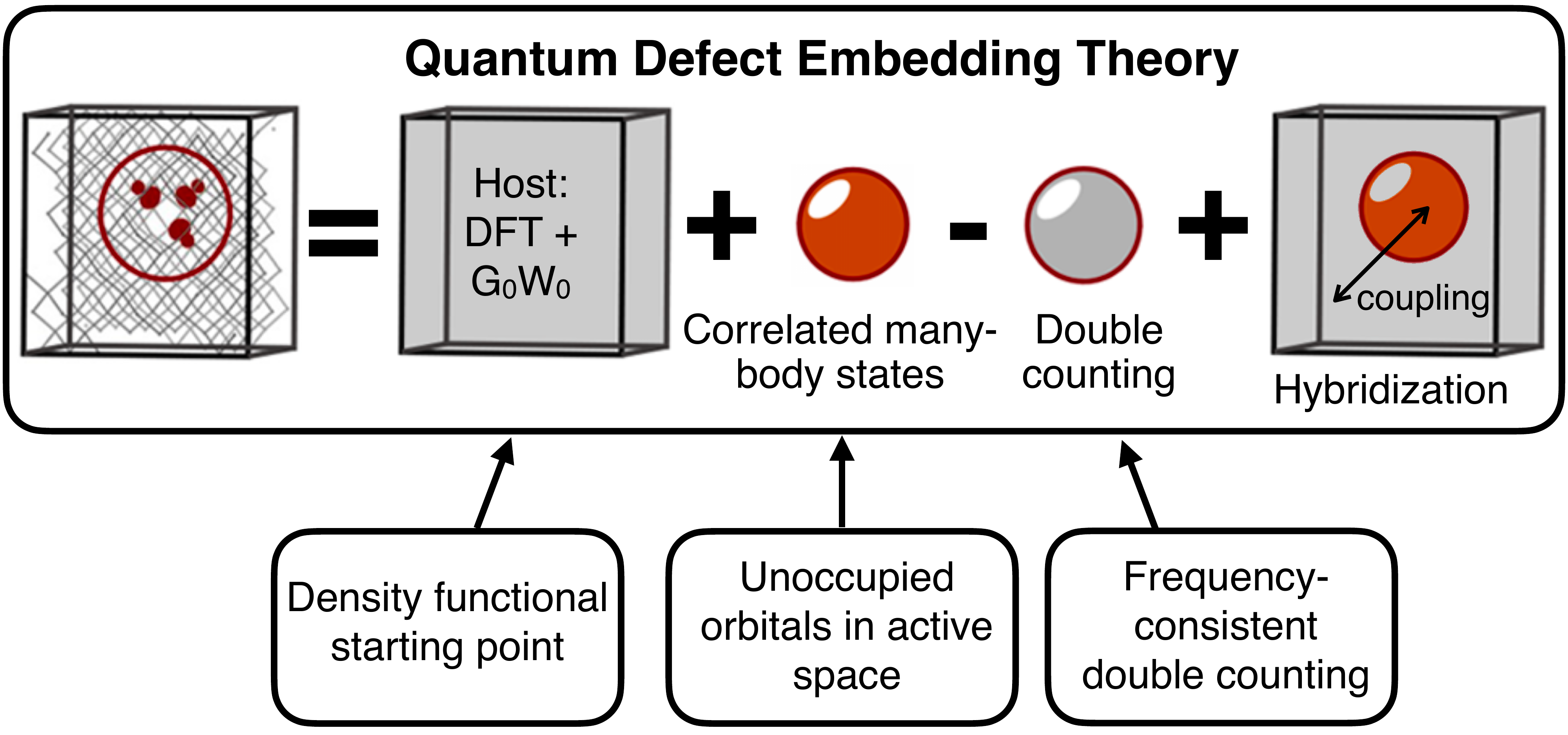}
\end{tocentry}

\begin{abstract}
Quantum defect embedding theory (QDET) is a many-body embedding method designed to describe condensed systems with correlated electrons localized within a given region of space, for example spin defects in semiconductors and insulators. Although the QDET approach has been successful in predicting the electronic properties of several point defects, several limitations of the method remain. In this work, we propose multiple advances to the QDET formalism. We derive a
double-counting correction that consistently treats the frequency dependence of the screened
Coulomb interaction, and we illustrate the effect of including unoccupied orbitals in the
active space. In addition, we propose a method to describe hybridization effects between the active
space and the environment, and we compare the results of several impurity solvers,
providing further insights into improving the reliability and applicability of the method. We present results for defects in diamond and for molecular qubits, including a detailed comparison with experiments.
\end{abstract}

\section{Introduction}

In the last decade, several many-body quantum embedding methods have been developed~\cite{Sun_2016_AccChemRes,Jones_2020_JACS,Vorwerk_2022_NCS,Bockstedte_2018_npjQM,Ma_2020_npjCM,Sheng_2022_JCTC,Muechler_2022_PRB} to describe the electronic structure of different regions of a complex system with varied levels of theory.
For example, the electronic states of point defects in semiconductors \cite{Xiong_2024_JACS} and insulators often require a high-level quantum chemistry treatment, as they are multi-reference states\cite{Ma_2020_PCCP,Thiering_npjCM_2019,Maze_2011_NJP}. However, their surrounding host may be described with lower-level methods, e.g. mean-field theories. In a similar fashion, one may consider a molecule or a nanoparticle with \CHANGES{multi-configurational} states interacting with a host surface or a matrix that may be described in an accurate manner with a mean-field theory.

In principle, one could describe the entirety of a complex system with many-electron methods such as quantum Monte Carlo \cite{QMC2001}, dynamical mean-field theory (DMFT)\cite{DMFT1996,DMFT2006}, or quantum chemistry methods \cite{Lischka_2018_ChemRev}, but the applicability of these approaches is limited to relatively small systems by their high computational cost, especially when one is interested in excited-state properties. Hence embedding methods provide a promising alternative, balancing cost and accuracy. Different embedding schemes have been proposed, based on partitioning the charge density\cite{Huang_2006_JCP,Huang_2011_JCP,Goodpaster_JCP_2014,Jacob_2014_WIREs,Genova_2014_JCP,Wen_2020_JCTC}, density matrix\cite{Knizia_2012_PRL,Pham_2020_JCTC,Wouters_2016_JCTC}, or Green's functions\cite{Kananenka_2015_PRB,NguyenLan_2016_JCTC,Aryasetiawan_2004_PRB,Zhu_2019_PRB,Miyake_2009_PRB,Romanova_JCP_2020,Ma_2020_npjCM,Ma_2020_PCCP,Ma_2021_JCTC,Sheng_2022_JCTC}, and utilizing various algorithms to embed a portion of a system into a larger environment. These include density matrix embedding theory (DMET)\cite{Knizia_2012_PRL,Knizia_2013_JCTC,Pham_2020_JCTC,Nusspickel_2022_PRX,Cui_2020_JCTC}, self-energy embedding theory (SEET)\cite{Kananenka_2015_PRB,NguyenLan_2016_JCTC,TranNguyen_2017_PRB,Zgid_2017_NJP}, and quantum defect embedding theory (QDET)\cite{Ma_2020_PCCP,Ma_2020_npjCM,Sheng_2022_JCTC,Vorwerk_2022_NCS}. These approaches differ in how they define the active space associated to the portion of the system treated with the high level of theory, and the choice of the low-level theory for describing the environment (e.g. Hartree-Fock versus density functional theory (DFT) or many-body perturbation theory), the treatment of the interaction between the active space and the environment (specifically, the treatment of double counting terms and the inclusion of hybridization between different portions of the system).

Here we focus on QDET and its application to describe point defects in solids, specifically promising defects to realize qubits for quantum sensing, computing, and communication technologies. \cite{Wolfowicz_2021_NRM, Schirhagl_2014_AnnuRevPhysChem,Barry_2020_RMP,Childress_2013_MRSBulletin,Weber_2010_PNAS,Waldherr_2014_Nature}. The scalability of QDET to large systems with up to 1,727 atoms \cite{Zhang_2025_arXiv} has been demonstrated in several papers, and shown to be instrumental in providing a description of realistic materials, with results directly comparable to experiments~\cite{Ma_2020_PCCP,Ma_2020_npjCM,Ma_2021_JCTC,Sheng_2022_JCTC,Muechler_2022_PRB,Otis_2025_JPCL}.

Previous applications of QDET have been successful in predicting the state ordering of various defect systems \cite{Ma_2020_npjCM, Ma_2020_PCCP,Ma_2021_JCTC, Sheng_2022_JCTC}, and in accurately estimating vertical excitation energies (VEEs) of several point defects of interest to quantum technologies\cite{Sheng_2022_JCTC,Wolfowicz_2021_NRM}. However, challenges remain in achieving the desired agreement with experiments when applying QDET to certain defects, including the moderately correlated neutral group IV vacancies in diamond \cite{Jin_2023_JCTC}.
These challenges may stem from several approximations adopted so far in the implementation of QDET, such as the absence of unoccupied orbitals in the active space, the neglect of hybridization between the active space and the environment and of the frequency dependence in the effective Hamiltonian, and the use of a non-self-consistent Green's function approach to describe the environment.

In this work, we propose several advances to the QDET formalism. First, we derive a double-counting correction that consistently 
treats the frequency dependence of the screened Coulomb interaction. 
We then illustrate the effect of including unoccupied orbitals in the active space, \CHANGES{and the effect of the choice of the exchange-correlation functional in the DFT calculations. Next, we} propose a method to describe hybridization effects between the active space and the environment. In addition, we compare the results of several impurity solvers, providing further insights into improving the reliability and applicability of QDET.

The rest of the paper is organized as follows. The Theory section includes a summary of the theoretical foundations of QDET, the 
 derivation of the double-counting term, and a method to account for hybridization effects.
 The Computational Setup section describes our computational workflow along with key strategies and algorithms used in our study of point defects. In the Results section, we analyze various defects in diamond, as well as molecular qubits, demonstrating the effects of the refined double-counting scheme, the inclusion of unoccupied orbitals in the active space, \CHANGES{the functional employed in the DFT calculations}, and of hybridization terms. In addition, we present a comparison of results obtained with four impurity solvers. We conclude the paper with a summary of our findings.

\section{Theory\label{sec:theory}}

QDET \cite{Ma_2020_npjCM,Ma_2020_PCCP,Ma_2021_JCTC,Sheng_2022_JCTC} is a Green's function embedding theory where the single-particle orbitals that describe a system of interacting electrons, $\{|\zeta_i\rangle\}$, obtained for instance by diagonalizing a Kohn-Sham (KS) Hamiltonian, are partitioned into two sets: active space (``$A$'') orbitals, for example localized orbitals associated to a point defect in a crystal, and environment (``$E$'') orbitals (that is, all remaining orbitals not included in the active space). \CHANGES{The electronic structure of the active space is described at a higher level of theory than that of the environment, due to the multi-configurational nature of the active space where the electronic states require a representation with several Slater determinants.}

The environment is described at the $G_0W_0$ level of theory, starting from DFT orbitals and the \CHANGES{correlated} active space is described using a quantum chemistry method, for example full configuration interaction (FCI).

Before delving into the theoretical derivation of QDET, we clarify our notation. 
We define the projector onto the active space as $f^A=\sum_{i\in A}|\zeta_i\rangle\langle\zeta_i|$, and similarly $f^E=\mathds{1}-f^A$ as the projector onto the environment.
Within a second quantization framework, we use \textit{subscripts} $A$, $E$, $AE$, and $EA$ to represent \textit{components} (submatrices or subtensors) of a given quantity defined in the entire system, i.e., $A\oplus E$. Specifically, subscripts $A$ and $E$ denote diagonal components in the active space and environment, respectively, and $AE,EA$ denote off-diagonal components. 
The matrix form of a two-indices quantity $M$ (e.g., one-body Hamiltonian $h$, Green's function $G$, self-energy $\Sigma$) defined in $A\oplus E$ (entire system) is:

\begin{equation}
    M = \begin{bmatrix}
        M_A & M_{AE} \\
        M_{EA} & M_E
    \end{bmatrix}\,,\label{eq:mat_M_components}
\end{equation}

We adopt the notation $Q_A$ for four-index quantities $Q$ (e.g., bare Coulomb interaction $v$, irreducible polarizability $P$, screened Coulomb interaction $W$) to denote the components of $Q$ where all four indices belong to $A$.

We use a \textit{superscript} $A$ to represent a \textit{projection} onto $A$ (for all orbital indices), without reduction of dimension. For a two-index quantity $M$, $M^A$ is given by

\begin{equation}
    M^A = f^A M f^A = \begin{bmatrix}
        M_A & 0 \\
        0 & 0
    \end{bmatrix},
\end{equation}
specifically:
\begin{equation}
    [M^A]_{ij} = \begin{cases}
    M_{ij}, & \mathrm{if}\,\,\,\, i,j \in A, \\
    0, &\mathrm{otherwise}.
    \end{cases}\label{eq:mat_M_proj_A}
\end{equation}
For a four-index quantity $Q$, $Q^A$ has the same dimension as $Q$ and is defined as:

\begin{equation}
    [Q^A]_{ijkl} = \begin{cases}
    Q_{ijkl}, & \mathrm{if}\,\,\,\, i,j,k,l \in A, \\
    0, & \mathrm{otherwise}.
    \end{cases}\label{eq:mat_Q_proj_A}
\end{equation}
We note that, in second quantization, $M^A$ and $Q^A$ have the same dimension of the $A\oplus E$ space, while $M_A$ and $Q_A$ have the same dimension of the $A$ subspace.

\subsection{Embedding scheme}\label{subsec:embed}

We consider the Dyson equations:
\begin{align}
  G^{-1}(\omega) &= \omega\mathds{1} - h - \Sigma(\omega), \label{eq:g_def} \\
  W^{-1}(\omega) &= v^{-1} - P(\omega) \label{eq:w_def},
\end{align}
where $G$ is the one-electron Green's function; $W$ is the 
screened Coulomb potential, $P(\omega)$ is the irreducible polarizability, $v$ is the bare Coulomb potential, and $h$ is the one-body, non-interacting part of the full electronic Hamiltonian. In \textit{ab initio} electronic calculations, $h$ expressed in first quantization is $h(\mathbf{r}) = -\frac{1}{2} \nabla^2 + v^{\mathrm{ion}}(\mathbf{r})$, where $v^{\mathrm{ion}}(\mathbf{r})$ is the electron-ion interaction.

We describe the active space at a higher level of theory (``HIGH'') than the environment (described at a ``LOW'' level of theory) and we assume that the self-energy in Eq.~\ref{eq:g_def} and irreducible polarizability in Eq.~\ref{eq:w_def} can be partitioned in the following way, \CHANGES{which is similar to the partitioning chosen in the $GW$+DMFT\cite{Ping_2002_PRB,Biermann_2003_PRL,Zhu_2021_PRX} and SEET methods\cite{Kananenka_2015_PRB,Zgid_2017_NJP,TranNguyen_2017_PRB}:}
\begin{align}
  \Sigma(\omega) &= \Sigma^{\mathrm{LOW}}(\omega) + \Sigma^{\mathrm{HIGH}}(\omega) -
  \Sigma^{\mathrm{DC}}(\omega), \label{eq:sigma_split}\\
  P(\omega) &= P^{\mathrm{LOW}}(\omega) + P^{\mathrm{HIGH}}(\omega) - P^{\mathrm{DC}}(\omega)
  ,\label{eq:p_split}
\end{align}
 where $\Sigma^\mathrm{DC}(\omega), P^\mathrm{DC}(\omega)$ are self-energy and irreducible polarizability terms that are double-counted in high- and
low-level descriptions of the active space $A$, respectively.
Note that in our embedding scheme, $\Sigma^\mathrm{HIGH}(\omega)$, $\Sigma^\mathrm{DC}(\omega)$, ${P}^\mathrm{HIGH}(\omega)$, and $P^\mathrm{DC}(\omega)$ have non-zero elements only in $A$, but they have the dimension of $A\oplus E$ in second quantization.

Inserting Eqs.~\ref{eq:sigma_split} and \ref{eq:p_split} into Eq.~\ref{eq:g_def}
and \ref{eq:w_def} yields:

\begin{align}
  G^{-1}(\omega) &= \omega\mathds{1} -h^R(\omega) - \Sigma^{\mathrm{HIGH}}(\omega), \label{eq:g_split} \\
  W^{-1}(\omega) &= \left[ W^R(\omega) \right]^{-1} - P^{\mathrm{HIGH}}(\omega) \label{eq:w_split},
\end{align}
where the frequency-dependent renormalized Hamiltonian $h^R(\omega)$ and partially
screened potential $W^R(\omega)$ are defined as
\begin{align}
  h^R(\omega) &= h + \Sigma^{\mathrm{LOW}}(\omega) -
  \Sigma^{\mathrm{DC}}(\omega), \label{eq:GRm1}\\
  \left[ W^R(\omega) \right]^{-1} &= v^{-1} - P^{\mathrm{LOW}}(\omega) + P^{\mathrm{DC}}(\omega). \label{eq:WRm1} 
\end{align}

Eqs.~\ref{eq:g_split} and \ref{eq:w_split} resemble Eqs.~\ref{eq:g_def} and
\ref{eq:w_def}. However, this resemblance does not imply that solving the electronic structure problem of the defect in the $A\oplus E$ space is equivalent to solving an effective impurity problem in $A$, using the high-level method, with a renormalized one-body Hamiltonian $h^R(\omega)$ and partially screened Coulomb potential $W^R(\omega)$.
The reason is that the quantities entering Eqs.~\ref{eq:g_split}-- \ref{eq:WRm1} are defined in the $A\oplus E$ space, not in $A$. 
Indeed, if we use Eq.~\ref{eq:g_split} to write the components of $G^{-1}(\omega)$ in the active space, we obtain
\begin{align}
  [G^{-1}(\omega)]_A = \omega \mathds{1}_A -h^R_A(\omega) - \Sigma^{\mathrm{HIGH}}_{A}(\omega) \,.\label{eq:g_split_proj_A}
\end{align}
However, to define a Green's function strictly for the active space, we must compute $G_A(\omega)$. The block matrix inversion formula shows that 
\begin{align}
    [G_A(\omega)]^{-1} = [G^{-1}(\omega)]_A - \Delta(\omega)\,,
\end{align}
where the hybridization term, $\Delta(\omega) = [G^{-1}(\omega)]_{AE} [[G^{-1}(\omega)]_{E}]^{-1} [G^{-1}(\omega)]_{EA}$, is 
\begin{align}
   \Delta(\omega) =  h^R_{AE}(\omega) [\omega \mathds{1}_E - h^R_E(\omega)]^{-1} h^R_{EA}(\omega) \,,
\end{align}
where we made use of the fact that $\Sigma^\mathrm{HIGH}(\omega)$ has nonzero components only in the active space.
The hybridization term vanishes if $h^R_{AE}(\omega)=h_{AE}+\Sigma^\mathrm{LOW}_{AE}(\omega)$ is negligible,
an assumption made in our previous formulation of QDET~\cite{Sheng_2022_JCTC}.
In Section~\ref{subsec:AH}, we show an approach to treat the hybridization term, but for now we assume that the hybridization term is negligible, i.e., $\Delta(\omega)\approx 0$; in this case, Eq.~\ref{eq:g_split_proj_A} can be simply rewritten as follows:

\begin{equation}
    [G_{A}(\omega)]^{-1} = \omega \mathds{1}_A -h^R_A(\omega) - \Sigma^{\mathrm{HIGH}}_{A}(\omega) \label{eq:g_split_A}.
\end{equation}

In addition, [see Section S1.1 in Supporting Information (SI)] Eq.~\ref{eq:w_split} can be rewritten as:
\begin{equation}
  [W_{A}(\omega)]^{-1} = \left[ W^R_{A}(\omega) \right]^{-1} - P^{\mathrm{HIGH}}_A(\omega) \label{eq:w_split_A}.
\end{equation}
\CHANGES{We note here that Eqs.~\ref{eq:WRm1} and \ref{eq:w_split_A} are in the same form as those adopted in the constrained random phase approximation (cRPA) proposed in Ref. \citenum{Aryasetiawan_2004_PRB}.}

Note that in Eqs.~\ref{eq:g_split_A}, \ref{eq:w_split_A}, $G_A(\omega)$ and $W_A(\omega)$ are defined exclusively in the active space. The similarity between Eqs.~\ref{eq:g_split_A}, \ref{eq:w_split_A} and Eqs.~\ref{eq:g_def}, \ref{eq:w_def} suggests that the electronic structure of the active space can be obtained by diagonalizing an effective Hamiltonian,

\begin{equation}\label{eq:def_hameff}
  H^{\mathrm{eff}} = \sum_{ij\in A} t^{\mathrm{eff}}_{ij}c^\dagger_i c_j +
  \frac{1}{2} \sum_{ijkl\in A} v^{\mathrm{eff}}_{ijkl}
  c^\dagger_i c^\dagger_j c_l c_k,
\end{equation}

In the following we derive expressions for the matrix elements $t^\mathrm{eff}$ and $v^\mathrm{eff}$. Comparing Eq.~\ref{eq:g_split_A} with Eq.~\ref{eq:g_def} we obtain an effective one-body interaction term $t^\mathrm{eff}$ from $h^R_A$, if we adopt the following approximation for $\Sigma^\mathrm{LOW}$ and $\Sigma^\mathrm{DC}$ that ensures that $t^\mathrm{eff}$ is frequency-independent and Hermitian:

\begin{equation}
\label{eq:teff_freq_dep}
\Sigma^\mathrm{LOW}_{ij} = \mathfrak{F}[\Sigma^\mathrm{LOW}(\omega)]_{ij}\,,\,\,\,\,
\Sigma^\mathrm{DC}_{ij} = \mathfrak{F}[\Sigma^\mathrm{DC}(\omega)]_{ij},
\end{equation}
where we have defined
\begin{equation}
\mathfrak{F}[M(\omega)]_{ij}\equiv \frac{1}{2}[\mathrm{Re}M_{ij}(\omega=E^{QP}_i)+\mathrm{Re}M_{ij}(\omega=E^{QP}_j)] \label{eq:qp_freq_approx}.
\end{equation}
Here $E^{QP}_i$ is the $G_0W_0$ quasiparticle energy of orbital $i$.
(This approximation is used in Ref.~\citenum{vanSchilfgaarde_2006_PRL} to implement the self-consistent $GW$ method.)
The effective one-body interaction is then

\begin{equation}
t^\mathrm{eff}\approx h_{A} + \Sigma^{\mathrm{LOW}}_{A} -
  \Sigma^{\mathrm{DC}}_{A}\,.\label{eq:def_teff}
\end{equation}

Comparing Eq.~\ref{eq:w_split_A} with Eq.~\ref{eq:w_def} we obtain an effective two-body interaction $v^\mathrm{eff}$:
\begin{equation}\label{eq:gen_veff}
  v^{\mathrm{eff}} \approx W^R_A(\omega=0),
\end{equation}
where $W^R(\omega)$ is approximated with a static value at zero frequency, a common choice in many-body perturbation theory (MBPT) \cite{Martin_2016_Book}.

Instead of solving the electronic structure of the entire system, in QDET we solve the effective Hamiltonian of Eq.~\ref{eq:def_hameff} to obtain the physical properties of the defect states, for example their vertical excitation energies.

The current implementation of QDET uses $G_0W_0$ as the low-level method, which allows us to define $\Sigma^\mathrm{LOW}(\omega)$ and $P^\mathrm{LOW}(\omega)$ in Eqs.~\ref{eq:sigma_split} and \ref{eq:p_split}. The self-energy $\Sigma^\mathrm{LOW}(\omega)$ is
\begin{equation}\label{eq:Sigma_LOW}
  \Sigma^\mathrm{LOW}(\omega) = v \rho_0 + \mathrm{i}\int\mathrm{d}\omega' G_0(\omega+\omega') W_0(\omega'),
\end{equation}
which is the sum of a Hartree term and the $G_0W_0$ exchange-correlation self-energy. The density $\rho_0$ and Green's function $G_0(\omega)$ are obtained from the KS Hamiltonian, 
\begin{equation}\label{eq:g0}
    G_0(\omega) = (\omega - H^{\mathrm{KS}})^{-1} = (\omega - h - v \rho_0-V^{xc})^{-1} \,,
\end{equation}
where $V^{xc}$ is the KS exchange-correlation potential.
The screened Coulomb potential is $W_0^{-1}(\omega) = v^{-1} - P_0(\omega)$, where the irreducible polarizability $P_0(\omega)$ is computed as:
\begin{equation}\label{eq:P0}
P_0(\omega)=-\mathrm{i}\int \mathrm{d}\omega' G_0(\omega+\omega')\times G_0(\omega')\,,
\end{equation}
where $\times$ denotes an outer product, and $P_0(\omega)$ is a four-orbital tensor, which in first quantization corresponds to $P_0(\mathbf{x},\mathbf{x'}; \omega)$ (used in Ref.~\citenum{Sheng_2022_JCTC}), and

\begin{equation}
    [P_0(\omega)]_{ijkl}= \int d\mathbf{x}d\mathbf{x'} \zeta_i(\mathbf{x})\zeta_k(\mathbf{x})P_0(\mathbf{x},\mathbf{x'}; \omega)\zeta_j(\mathbf{x'})\zeta_l(\mathbf{x'})\,.
\end{equation}
The polarizability $P^\mathrm{LOW}(\omega)$ is
\begin{equation}
    P^\mathrm{LOW}(\omega)=P_0(\omega)\,.\label{eq:P_LOW}
\end{equation}

\subsection{Double counting}\label{subsec:DC}

To derive the double-counting terms $\Sigma^\mathrm{DC}$ and $P^\mathrm{DC}$, we require that the terms $\Sigma^\mathrm{DC}(\omega)$ and $P^\mathrm{DC}(\omega)$ defined in the active space coincide with the self-energy and irreducible polarizability obtained from the low-level theory (DFT+$G_0W_0$) when using the effective Hamiltonian $H=H^\mathrm{eff}$,

\begin{align}
  \Sigma^\mathrm{DC}_A(\omega) 
  &=v^\mathrm{eff}\rho_{0,A} + \mathrm{i}\int\mathrm{d}\omega' G_{0,A}(\omega+\omega') W^\mathrm{eff}_A(\omega')\,,\label{eq:dc_sigma} \\
  P^\mathrm{DC}_A(\omega) &= P_{0,A}(\omega)
  = -\mathrm{i}\int \mathrm{d}\omega' G_{0,A}(\omega+\omega')\times G_{0,A}(\omega') \,.\label{eq:dc_P}
\end{align}
These equations are obtained by replacing the ``full-system'' quantities with ``effective'' quantities in Eqs.~\ref{eq:Sigma_LOW} and \ref{eq:P_LOW}: 
The bare Coulomb potential $v$ is replaced by the effective Coulomb potential $v^\mathrm{eff}$. The density, Green's function, and irreducible polarizability are replaced by the DFT values in the active space: $\rho_{0,A}$, 
$G_{0,A}(\omega)$, and
$P_{0,A}(\omega)$, respectively; this is an assumption due to the absence of self-consistency.
Lastly, $W_0(\omega)$ is replaced by the effective screened Coulomb potential $W^\mathrm{eff}(\omega)$ (see Section S1.2 in SI for the derivation),

\begin{equation}
W^\mathrm{eff}(\omega) = [v^{-1} - P_0^R(\omega=0) - P_0^A(\omega)]^{-1},\label{eq:dc_W}
\end{equation}
where we have defined the ``reduced'' irreducible polarizability $P_0^R(\omega)=P_0(\omega)-P_0^A(\omega)$.
Note the use of the superscript $A$ defined at the beginning of the section.

Now, combining Eqs.~\ref{eq:def_teff}, \ref{eq:gen_veff}, and \ref{eq:dc_sigma}, we obtain $t^\mathrm{eff} = H^\mathrm{KS}_{A} - t^\mathrm{DC}_{A} $, in which

\begin{align}
    t^\mathrm{DC} =& 
    V^\mathrm{xc} + W^R(\omega=0)\rho_0^A \nonumber\\ 
    &- \mathrm{i}\int\mathrm{d}\omega' G_0(\omega+\omega')W_0(\omega')\nonumber\\
    &+ \mathrm{i}\int\mathrm{d}\omega' G_0^A(\omega+\omega')W^\mathrm{eff}(\omega')\,.
    \label{eq:tdc_final}
\end{align}
The last two terms of the right hand side of Eq.~\ref{eq:tdc_final} are $\Sigma^\mathrm{LOW}(\omega)$ and $\Sigma^\mathrm{DC}(\omega)$, respectively. Following the same procedure as in Eq.~\ref{eq:teff_freq_dep}, we evaluate $\Sigma^\mathrm{LOW}(\omega)$ and $\Sigma^\mathrm{DC}(\omega)$ at the quasi-particle energy [$\mathfrak{F}(\cdot)$], thus removing any frequency dependence and ensuring the effective Hamiltonian is Hermitian. We denote this double-counting scheme as ``refined double counting'' (DC2025).

The difference between Eq.~\ref{eq:tdc_final} here and Eq. 26 in Ref.~\citenum{Sheng_2022_JCTC} (which we denote DC2022) is due to $W^\mathrm{eff}(\omega)$: in DC2022, $W^\mathrm{eff}(\omega)$ = $W_0(\omega)$, since Eq.~\ref{eq:dc_W} was derived without explicitly considering the frequency dependence of each term (in other words, the same frequency was used for each frequency-dependent term):
\begin{equation}
W^\mathrm{eff} = [v^{-1} - P_0^R - P_0^A]^{-1} = W_0 \label{eq:dc2022},
\end{equation}
which leads to a cancellation of the last two terms in Eq.~\ref{eq:tdc_final}.
In DC2025, the frequency dependence of each term is explicitly taken into account, 
which 
ensures that the double counting term entering $t^\mathrm{eff}$ is correct, despite the fact that $v^\mathrm{eff}$ is frequency independent.

\subsection{Approximate hybridization}\label{subsec:AH}

We now consider the case where $h^R_{AE}(\omega)$ is not negligible, and hence the hybridization term does not vanish, i.e. $\Delta(\omega)\neq0$. 
Using Eq.~\ref{eq:teff_freq_dep}, we obtain the following ``sum-of-poles" approximation for $\Delta(\omega)$:
\begin{align}
  \Delta_{ij}(\omega) = \sum_{b \in \mathrm{poles}} \Gamma_{ib} \frac{1}{\omega-\epsilon_b} \Gamma_{bj}, \;\;\;\; \,\,\,\, i,j \in A \label{eq:Delta_2}
\end{align}
The energies $\epsilon_b$ and the terms $\Gamma_{ib}$ can be obtained for instance via a diagonalization of $h^R_E$.

With the hybridization present, Eq.~\ref{eq:g_split_A} is rewritten as

\begin{equation}
[G_{A}(\omega)]^{-1} = \omega \mathds{1}_A -h^R_A(\omega) -\Delta(\omega) - \Sigma^{\mathrm{HIGH}}_{A}(\omega)\,.\label{eq:G_hyb}
\end{equation}

We construct an auxiliary Hamiltonian defined in the Hilbert space formed by the active space orbitals plus a given set of ``bath orbitals''; the Hamiltonian is such that the interactions between active space and bath or within the bath are strictly one-body:

\begin{equation}
    H^\mathrm{aux} = \sum_{ij} t^\mathrm{eff}_{ij} c^\dagger_i c_j + \frac{1}{2} \sum_{ijkl} v^\mathrm{eff}_{ijkl} c^\dagger_i c^\dagger_j c_l c_k + \left(\sum_{ib} \Gamma_{ib}c^\dagger_i c_b + h.c. \right) + \sum_{b}\epsilon_b c^\dagger_b c_b,\label{eq:Haux}
\end{equation}
where $i,j,k,l$ loop over active space orbitals and $b$ loops over bath orbitals. 
This Hamiltonian is defined on the space of $A\oplus B$, where $B$ is the space of the bath orbitals. 
The Green's function of this Hamiltonian has the same form as
in Eq.~\ref{eq:G_hyb}, with the hybridization term in $A$ expressed exactly as in Eq.~\ref{eq:Delta_2}\cite{Zgid_2011_JCP}. 
Hence we can solve the effective Hamiltonian with hybridization by diagonalizing the auxiliary Hamiltonian; the effective Hamiltonian with hybridization and the auxiliary Hamiltonian yield identical Green's functions. The advantage of the auxiliary Hamiltonian is that it is frequency-independent and can be solved with a chemistry impurity solver.

Note that the hybridization scheme adopted here is approximate for several reasons. First, we have assumed that $\Sigma^\mathrm{LOW}$ is frequency independent, using the approximation in Eq.~\ref{eq:qp_freq_approx}.
When defining $t^\mathrm{eff}$, frequency-dependent terms entering the exact expressions of $\Sigma^\mathrm{LOW}(\omega)$ and $\Sigma^\mathrm{DC}(\omega)$ may cancel each other, thus rendering $t^\mathrm{eff}$ weakly dependent on frequency. However, we do not have any cancellation effects to rely on in the case of the hybridization.
Second, writing $\Gamma$ in a second-quantized matrix form implies that this quantity is defined, in practice, on a finite basis set. 
When the basis is truncated, $f^A+f^E=\mathds{1}$ might not be satisfied.
Alternatively, one may use $f^E=\mathds{1}-f^A$, but in this case 
we cannot any longer diagonalize $h^R_E$ and
we would need to resort to a highly nonlinear fitting algorithm \cite{Zgid_2011_JCP} to determine the $\Gamma_{ib}$ and $\epsilon_b$ parameters of the auxiliary Hamiltonian. This fitting procedure may not be precise and in practice it is not feasible for the system sizes we are interested in here.

We also note the lack of self-consistency in the determination of the hybridization term: unlike DMET or SEET in which the hybridization term may be iteratively updated, either through a self-consistent low-level method, or the definition of multiple fragments, in QDET we cannot at present update $\Delta(\omega)$, since only the orbitals of a single defect are in the active space, and the low-level method is $G_0W_0$. Future improvements of the method could explore frequency dependence or a self-consistent low-level solver.

\section{Computational Setup\label{sec:comp_setup}}

\begin{figure}
\includegraphics[width=1.1\linewidth]{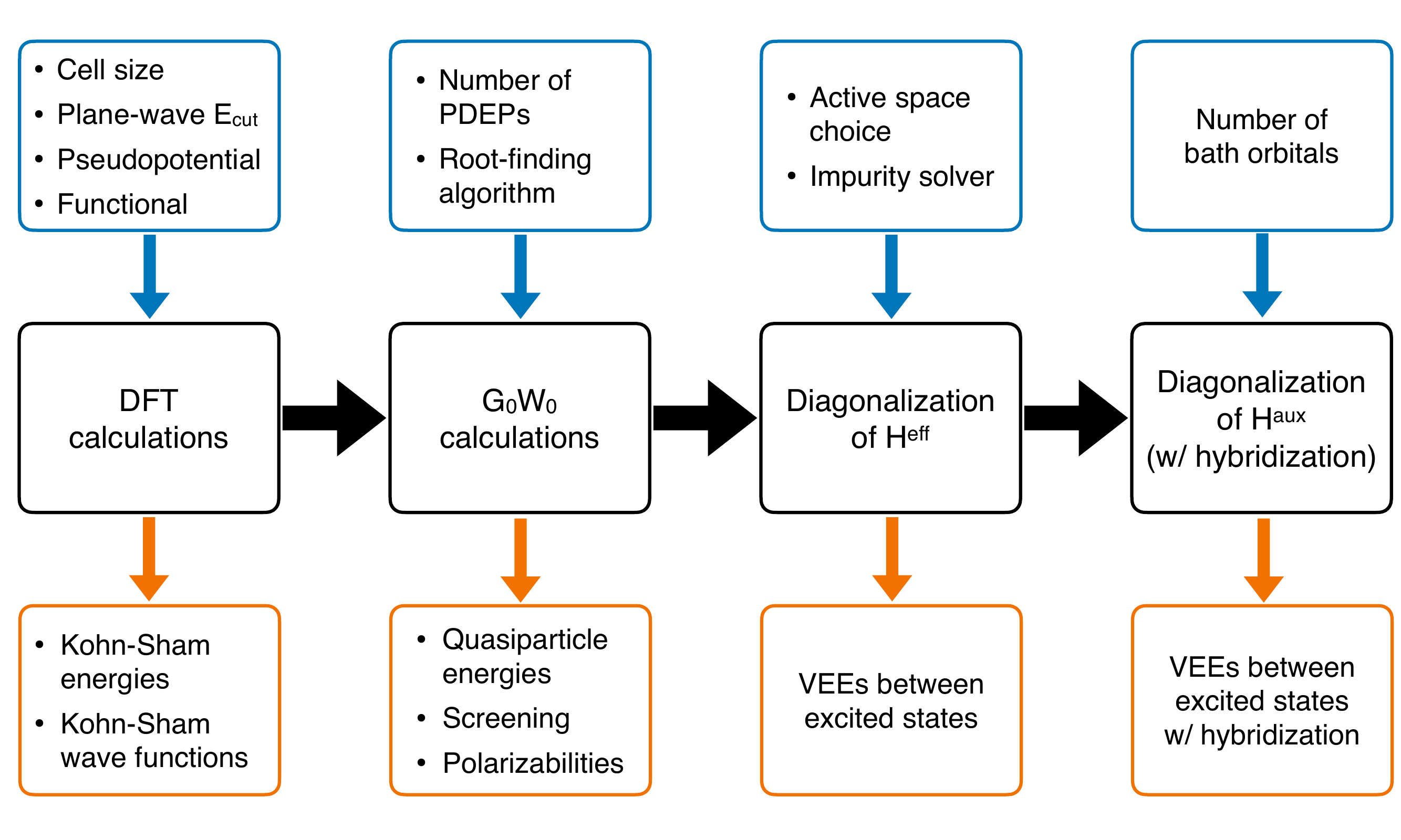}
\caption{\label{fig:workflow}The workflow of QDET calculations carried out in this work, including the input (top row) and output (bottom row) of each step.
The effective and auxiliary Hamiltonians $H^\mathrm{eff}$ and $H^\mathrm{aux}$ are defined in Eq.~\ref{eq:def_hameff} and ~\ref{eq:Haux}, respectively.
VEE denotes vertical ionization energies and PDEP denotes the projective dielectric eigenpotentials (see Section~\ref{subsec:PDEP}). Screening denotes the screened Coulomb interactions.}
\end{figure}

The computational workflow adopted in our study is summarized in Fig.~\ref{fig:workflow}. Using atomic positions optimized with unrestricted DFT calculations, we first perform a restricted DFT calculation with preset occupation numbers as shown in Figs. \ref{fig:newDC_NV_511}(b) and \ref{fig:newDC_groupIV_511}(b). The DFT calculations used the Quantum ESPRESSO (QE) \cite{QE-2009,QE-2017,QE-GPU} code, the PBE functional \cite{PBE}, and the SG15 PBE norm-conserving pseudopotentials \cite{SG15}. \CHANGES{Calculations using the dielectric dependent hybrid (DDH) functional \cite{Skone_2014_PRB,Skone_2016_PRB} are presented in Section~\ref{subsec:ddh}.} A plane wave kinetic energy cutoff of 60 Ry is used for all diamond defects and for the Cr(\textit{o}-tolyl)$_4$ molecular qubit. The Brillouin zone of the supercell is sampled with the $\Gamma$-point only.
Starting from KS eigenvalues and eigenfunctions, we perform a $G_0W_0$ calculation with the WEST code \cite{Govoni_2015_JCTC,Yu_2022_JCTC}. We then choose an active space, for which an effective Hamiltonian is generated following the method described in Section~\ref{sec:theory}.
This Hamiltonian is diagonalized with one of the impurity solvers described in Section~\ref{subsec:imp_solver} to obtain ground and excited-state energies of many-body electronic states. 
To account for hybridization, we use the $G_0W_0$ self-energies to compute a finite number of bath orbitals and define the auxiliary Hamiltonian, as described in Section~\ref{sec:theory}. We then use an impurity solver to diagonalize this Hamiltonian as well, thus obtaining the VEEs of the system with an approximate hybridization term.

\CHANGES{The computational cost of a QDET calculation is determined by: (i) the evaluation of the dielectric screening, (ii) the construction of the QDET Hamiltonian, and (iii) the diagonalization of the Hamiltonian with a chosen impurity solver. The cost of computing the dielectric screening is independent of the size of the active space and scales as $O(N_\mathrm{tot}^4)$, where $N_\mathrm{tot}$ is the total number of electrons in the system.
The cost of constructing the QDET Hamiltonian is proportional to the number of self-energy matrix elements and scales as $O(N_\mathrm{tot}^3 M^2)$, where $M$ is the total number of active space orbitals (occupied and unoccupied).
The cost of diagonalizing the Hamiltonian depends on the choice of the number of orbitals in the active space and on the choice of the impurity solver, discussed in Section~\ref{subsec:imp_solver}.
}

We describe next the details of our computational procedure, including convergence tests.

\subsection{Convergence of G$_0$W$_0$ calculations\label{subsec:PDEP}}

In the WEST code, a low-rank representation of the screened Coulomb interaction $W_0$ is used, as obtained from a projected eigen-decomposition of the dielectric matrix \cite{Wilson_2008_PRB,Wilson_2009_PRB}. The number of eigenvectors (or eigenpotentials) of the dielectric matrix (which we refer to as ``PDEPs'') entering the low-rank representation of $W_0$ controls the numerical accuracy of the calculations. We have verified (see Section S2 in SI) that to converge the VEEs computed here within 10 meV, $G_0W_0$ calculations require a number of PDEPs equal to two to three times the number of electrons in the supercell. In all calculations reported below, we set the number of PDEPs to be three times the number of electrons.

\subsection{Selection of active space}

Several different criteria may be adopted to select the single-particle orbitals in the active space. We consider two of them \CHANGES{(and briefly discuss an additional one in the SI)}:
(1) Compute the localization factor $L_{\Omega}(\zeta_i) = \int_{\textbf{r} \subseteq \Omega} | \zeta_i(\textbf{r}) |^2 d\textbf{r}$ within a chosen volume $\Omega$ for each KS orbital $\zeta_i$, then select the $N$ orbitals with the largest $L_{\Omega}$~\cite{Sheng_2022_JCTC}. The convergence of the procedure may be controlled by increasing $N$.
(2) Choose first the minimum number of single-particle orbitals required to describe the excited states~\cite{Ma_2020_PCCP} of the defect (``minimum model'', ``MM''); then augment this minimum model by incrementally adding occupied KS orbitals with energy close to the valence band maximum (VBM) and, when computationally feasible, unoccupied KS orbitals close to the conduction band minimum (CBM).

We compared the convergence of criteria (1) and (2) for the negatively charged nitrogen-vacancy center (NV$^-$) and the neutral silicon vacancy (SiV$^0$) in diamond.
We integrated $L_{\Omega}$ within a sphere of radius 1.54 \AA\
(equal to the distance between first neighbors in diamond), centered at the carbon vacancy for NV$^-$ and at the Si atom for SiV$^0$.
 Our results %
(see Section S3 in SI) show that (2) leads to a smoother convergence of the VEEs with respect to the size of the active space compared to (1).
In addition, the definition of the localization factor in (1) requires the choice of an integration radius and thus careful examination of the specific defect under consideration. On the other hand, the minimum model chosen in (2) has an unambiguous and physically motivated definition, to which one can increasingly add states with energy close to the band edges.

Therefore, we chose to carry out all of our calculations with criterion (2), which we denote as a minimum model plus KS energy (MM+KSE) approach.
Details on the minimum model and the energy threshold chosen for each defect are discussed in Section~\ref{sec:results}.

We note that most of our computations include only (fully and partially) occupied KS orbitals. This is primarily due to the prohibitive cost of diagonalizing the effective Hamiltonian when unoccupied orbitals are included. Additionally, the presence of unoccupied orbitals in the active space may give rise to solutions that correspond to bound excitons. These solutions are physical, however they are found at an incorrect energy, since they involve not only localized orbitals, but both localized and delocalized orbitals that the QDET scheme cannot accurately describe. In Ref.~\citenum{Jin_2023_JCTC}, we discussed the requirements to accurately describe a bound exciton for the SiV$^0$ defect, in terms of cell sizes and finite size scaling. We showed that convergence may be attained only with more than 1,700 atoms in the supercell.
Note that localized and delocalized orbitals exhibit different behaviors as a function of the supercell size.

Solutions corresponding to bound excitons
can be identified and do not interfere with the calculation of defects' excited states of interest; however, in practice, the bound-exciton solutions lead to an increase in the number of excited states to be computed by the impurity solver, and thus to an increase in the computational cost of the diagonalization.

Nevertheless, in Section~\ref{sec:results}, we present some calculations obtained with active spaces including unoccupied orbitals, when feasible. We show that, for neutral group IV vacancies in diamond and the Cr(\textit{o}-tolyl)$_4$ molecular qubit, excluding unoccupied orbitals from the active space only introduces small errors (on the order of 0.02 eV) in the computed VEEs. For the NV center, this effect is larger, on the order of 0.1 eV.

\subsection{Selection of bath orbitals}\label{subsec:bath_orb}

As described in Section~\ref{sec:theory}, the expression of the hybridization $\Delta(\omega)$ is obtained as a sum-of-poles decomposition. In principle, the number of poles should be equal to the number of orbitals in the environment, but it is unfeasible to include all of them in the calculation. We sort the poles according to their ``contribution'' $\mathcal{S}_b$ to hybridization, then select the highest ones until the total contribution reaches a given threshold $\sum_b \mathcal{S}_b \geq \mathcal{T}$, using the formula: %
\begin{equation}
\label{eq:bath_orb_contribution}
\mathcal{S}_b=\frac{\sum_i|\Gamma_{ib}|^2}{\sum_{ib'}|\Gamma_{ib'}|^2},
\end{equation}
where $\Gamma$ is defined in Section~\ref{subsec:AH}, $i$ loops over active space orbitals, and $b$, $b'$ loop over poles.

The selected poles define the bath orbitals that we use to construct the auxiliary Hamiltonian. The accuracy of our selection criterion and convergence as a function of the number of poles are presented in Section S4 of SI.

Additionally, the number of electrons in the auxiliary system (active-plus-bath) needs to be carefully set. 
To do so, we first select the number of electrons by filling all bath orbitals with pole frequency lower than the Fermi level; the rest of the bath orbitals are initially empty. We then adjust the number of filled orbitals to obtain the correct number of active-space electrons in the ground state of the auxiliary system, computed, for example, via a one-body density matrix.

\CHANGES{
An alternative approach would be to directly include the bath orbitals in the active space. 
While technically feasible, this approach is not yet supported by the current QDET framework, which is formulated in terms of KS orbitals, and uses quasi-particle energies as frequencies of the self-energies (Eq.~\ref{eq:qp_freq_approx}). Bath orbitals are linear combinations of KS orbitals, for which such frequency approximation is ill-defined. 
}

\subsection{Choice of impurity solver}\label{subsec:imp_solver}

We used four different impurity solvers to diagonalize the auxiliary Hamiltonian, and we compare their efficiency and accuracy in Section~\ref{sec:results}.

\textbf{Full Configuration Interaction (FCI)}. If neither approximate hybridization nor unoccupied orbitals are considered, most of the impurity problems investigated here can be solved using FCI. The Hamiltonian is then either exactly diagonalized (ED), or iteratively diagonalized with a Lanczos solver, to obtain the states at the desired energies. We use PySCF \cite{PySCF-2018,PySCF-2020} to perform FCI calculations and symmetry analyses. 
\CHANGES{
Using a Lanczos solver, the diagonalization cost scales as $O(LN_\mathrm{conf}^2)$, where $N_\mathrm{conf}$ is the number of configurations in the FCI space, and $L$ is the number of many-body states to be computed. When neither unoccupied orbitals nor bath orbitals are included in the active space, $O(LN_\mathrm{conf}^2)\approx O(LM_o^4)$, where $M_o$ is the number of occupied orbitals in the active space. }

When unoccupied or bath orbitals are included, 
\CHANGES{the number of configurations in the FCI space and the corresponding computational cost both become factorial, rendering FCI calculations unfeasible in most cases.} Hence, we resort to alternative solvers, briefly described below.

\textbf{Selected Configuration Interaction (CI)}. One of these solvers uses the selected CI algorithm \cite{Sharma_2017_JCTC,Holmes_2016_JCTC,Tubman_2016_JCP,Evangelista_2014_JCP,Giner_2013_CJC}, implemented in PySCF. Selected CI is particularly efficient when the Hamiltonian matrix elements involving unoccupied orbitals are small relative to those between occupied states. 
Note that the implementation of selected CI in PySCF does not support the diagonalization of Hamiltonians of the form of the auxiliary Hamiltonian. 
\CHANGES{
The computational cost of selected CI depends on the implementation and the sparsity of the Hamiltonian. For the systems tested here, we observed that although the computational cost of selected CI is on average much smaller than that of FCI, it is still super-polynomial as a function of the number of active space orbitals.
}

\textbf{\CHANGES{Multi-reference Configuration Interaction Single and Doubles (MR-CISD/CIS(D))}}. \CHANGES{An alternative solver to FCI is the MR-CISD algorithm. The CI space of this solver consists of all single and double excitations from a set of reference determinants, which are constructed by considering all possible occupations of the partially occupied orbitals. For example, in the case of the NV$^-$ center in diamond, the reference space comprises all configurations formed by distributing two electrons among the two degenerate, partially occupied $e$ orbitals. }

In the case of the auxiliary Hamiltonian defined to account for hybridization, two-body interactions are considered only within the active space, while the active-bath and bath-bath interactions are purely one-body. Hence, double excitations outside the active space can be neglected in the CI procedure. \CHANGES{We denote the MR-CISD algorithm with this simplification as MR-CIS(D), to emphasize that double excitations are not fully considered. This simplification} significantly reduces the number of configurations and computational cost. 

An exact diagonalization or Lanczos solver is then employed to solve the Hamiltonian.
\CHANGES{
The computational cost of MR-CISD scales as $O(LN_\mathrm{conf}^2)\approx O(LM_o^4M_u^4)$, where $M_u$ is the number of unoccupied orbitals. 
Unlike selected CI, the computational cost of MR-CISD does not depend on the sparsity of the Hamiltonian. Therefore, although MR-CISD can be more expensive than selected CI in the case of a sparse Hamiltonian, its computational cost is on average smaller than that of selected CI for an arbitrary Hamiltonian.
}

\textbf{Auxiliary Field Quantum Monte Carlo (AFQMC)}. Auxiliary-field quantum Monte Carlo \cite{Zhang_1997_PRB,Zhang_2003_PRL} \CHANGES{can be used to validate MR-CISD/CIS(D) results when neither FCI nor selected CI is computationally feasible. It is} an imaginary-time projector Monte Carlo method that has been shown to yield excellent results for ground state properties\cite{Williams_2020_PRX,Motta_2018_WIREs,Chen_2023_PRB}; recently it has been applied to study excited states as well\cite{Ma_2013_NJP,Shee_2019_JCTC}. Here we use the constraint-path generic-basis AFQMC with multi-determinant trial wave functions (GAFQMC-MD)\cite{AlSaidi_2006_JCP,Motta_2018_WIREs}. This approach to computing excited states is especially suitable for our defect problems, where each many-body state has a well-defined symmetry. The selection of a trial wave function of a given symmetry to guide the projection automatically excludes all states with incompatible symmetries, making the algorithm particularly efficient. %
In our calculations, we selected the trial wave functions as the wave functions obtained using the aforementioned MR-CISD/CIS(D) solver.

The computational complexity of AFQMC grows as $O(L M^2 N^2)$ where $L$ is the number of states, $M$ is the number of orbitals, and $N$ is the number of electrons \cite{Shi_2021_JCP}. Hence AFQMC is a better choice than CI-based methods for large systems. As a near-exact method, it is less accurate than the (almost) exact solvers like FCI or selected CI, but it most often yields results within chemical accuracy (0.04 eV)\cite{Williams_2020_PRX}.

\section{Results\label{sec:results}}
\subsection{Double counting corrections}\label{subsec:dc_compare}

\begin{figure}
\includegraphics[width=0.8\linewidth]{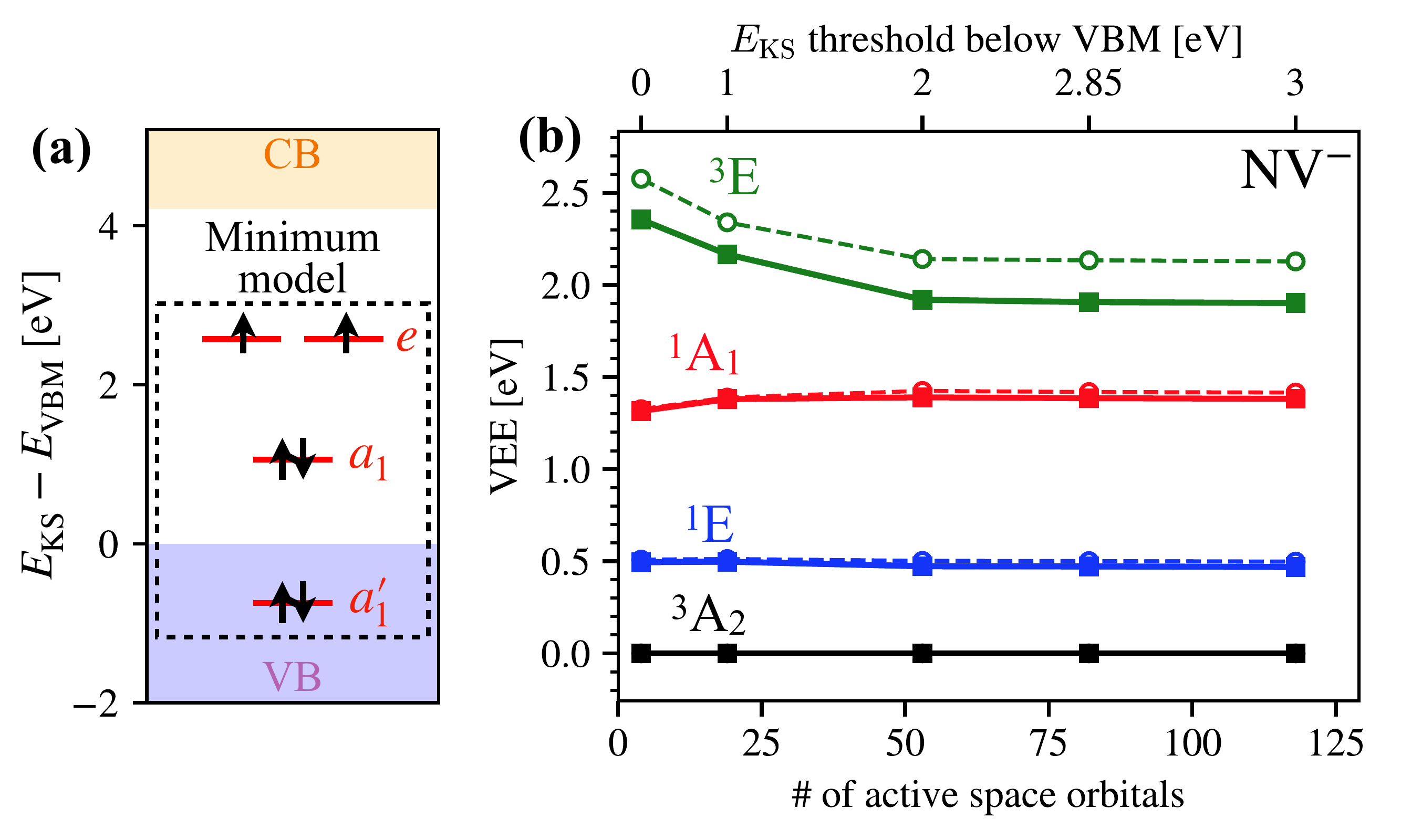}
\caption{\label{fig:newDC_NV_511}
(a) Minimum model included in the active space of the NV center in diamond and (b) comparison of computed many-body states using the DC2022 (dashed lines) and DC2025 (solid lines) double-counting terms, where the vertical excitation energies (VEEs) are reported as a function of the number of orbitals in the active space \CHANGES{and as a function of the Kohn-Sham energy ($E_\mathrm{KS}$) threshold below VBM}. See text for the definition of DC2022 and DC2025. Calculations are carried out in a 511-atom supercell, with the PBE functional to obtain Kohn-Sham wave functions.}
\end{figure}

We first study the NV$^-$ in diamond in 215-atom and 511-atom supercells, starting from a minimum model of 4 orbitals [$e,a_1,a_1'$ in the active space, see Fig.~\ref{fig:newDC_NV_511}(b)], and we compare results obtained with DC2022 and DC2025. In this section, unoccupied KS orbitals are not included in the active space. Fig.~\ref{fig:newDC_NV_511}(c) shows our results for the 511-atom supercell. 
Compared to DC2022, using DC2025 leads to a decrease in the VEE of the $^3E$ state, irrespective of the active space composition. The converged VEE for this state in the 215-atom supercell is 1.939 eV, and in the 511-atom supercell is 1.902 eV, in apparent worse agreement with experiments (2.18 eV \cite{Davies_1976_PRSA}) than the results obtained with DC2022. \CHANGES{However, we will show below that the agreement with the experiment is restored when effects from unoccupied orbitals and hybrid functional are included. }

\begin{figure}
\includegraphics[width=0.8\linewidth]{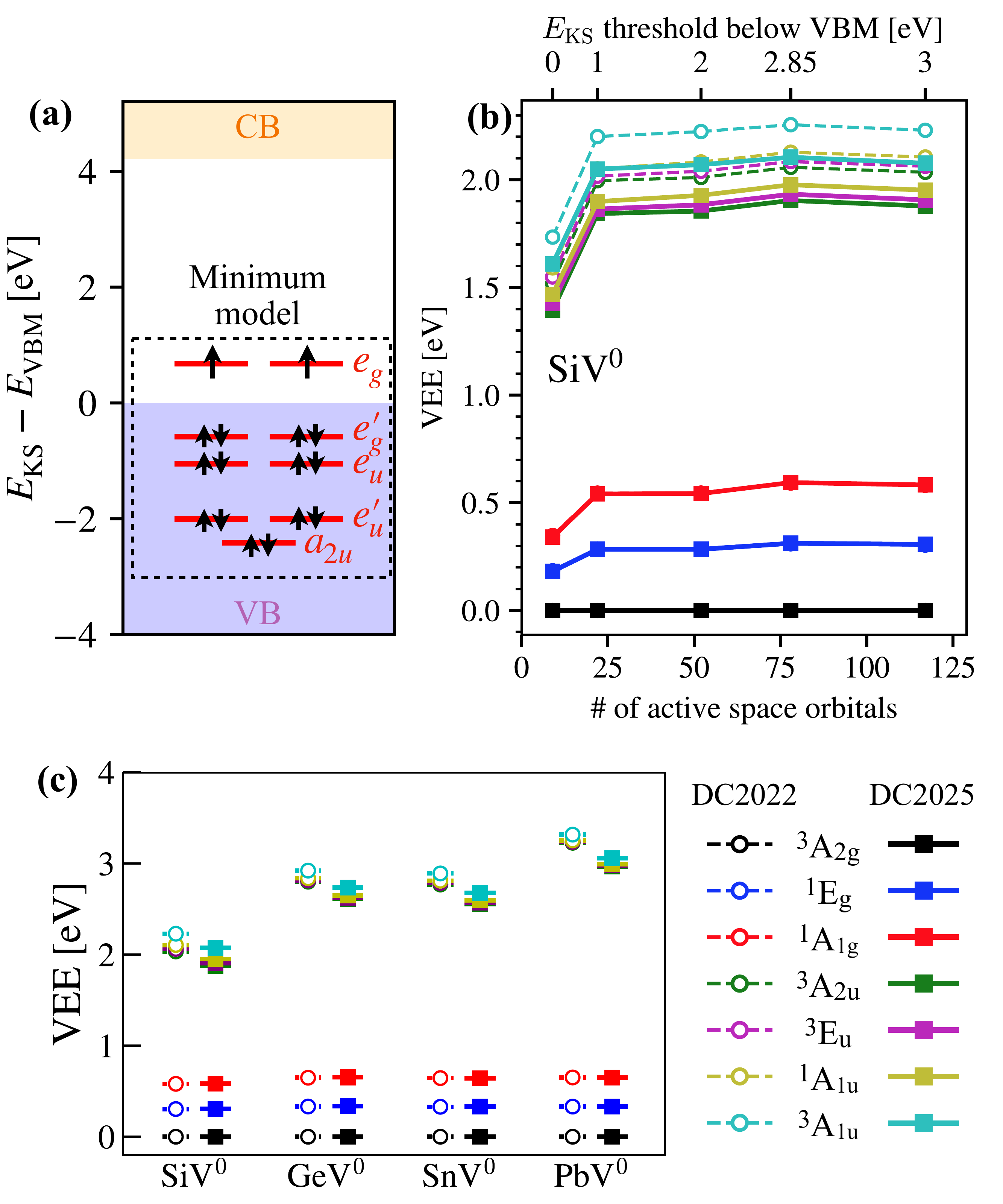}
\caption{\label{fig:newDC_groupIV_511} (a) Minimum model included in the active space of the SiV$^0$ center in diamond; (b) comparison of computed many-body states using the DC2022 (dashed lines) and DC2025 (solid lines) double-counting terms, where the vertical excitation energies (VEEs) are reported as a function of the number of orbitals in the active space \CHANGES{and as a function of the Kohn-Sham energy ($E_\mathrm{KS}$) threshold below VBM} (see text for the definition of DC2022 and DC2025); 
(c) comparison of VEE obtained with DC2022 and DC2025 for all group IV neutral vacancy defects, computed with the largest active space considered in this work (minimum model plus occupied orbitals with KS energies 3 eV below the VBM). Calculations are carried out in a 511-atom supercell, with the PBE functional to obtain Kohn-Sham wave functions.}
\end{figure}

We also carried out the same comparison for neutral group IV vacancies in diamond, starting from a 9-orbital minimum model [$e_g,e_u,e'_g,e'_u,a_{2u}$, Fig.~\ref{fig:newDC_groupIV_511}(b)], and gradually adding occupied KS orbitals with the largest $\lambda_i$. Results obtained with a 511-atom supercell for SiV$^0$ are given in Fig.~\ref{fig:newDC_groupIV_511}(c). We observe again a decrease in the VEEs of the higher excited states of about 0.15 eV, when using DC2025; similar to the NV$^-$ center, the decrease is almost the same, irrespective of the choice of the active space. In the case of SiV$^0$, we obtain results in closer agreement with experiments (1.5--1.6 eV \cite{DHaenensJohansson_2011_PRB}) than with DC2022; we show below that the inclusion of unoccupied orbitals in the active space does not substantially affect the results obtained here.
Other group IV defects show a similar convergence trend as a function of the active space size (see Section S5 in SI). In Fig.~\ref{fig:newDC_groupIV_511}(d), we compare the results with DC2022 and DC2025 double-counting terms for the largest active space only.
For the germanium, tin, and lead vacancies (GeV$^0$, SnV$^0$, and PbV$^0$), we observe a decrease in the highest excited-state VEEs, when using DC2025, with variations of 0.2--0.25 eV.

\subsection{Effect of unoccupied orbitals}

In this section, we include several unoccupied orbitals in the active space and investigate their effect on computed VEEs. Starting from a reference active space that contains only occupied orbitals, we add, one at a time, several groups of low-lying unoccupied orbitals close in (KS) energy. The addition of the combination of all sets of orbitals at the same time is not computationally feasible. If we assume that the effect of adding several sets of unoccupied orbitals, relative to the reference active space, is additive, then we can estimate their total effect. However, additivity only holds for sets of orbitals that do not hybridize with each other, hence our estimate is necessarily approximate. 

\begin{figure}
\includegraphics[width=0.8\linewidth]{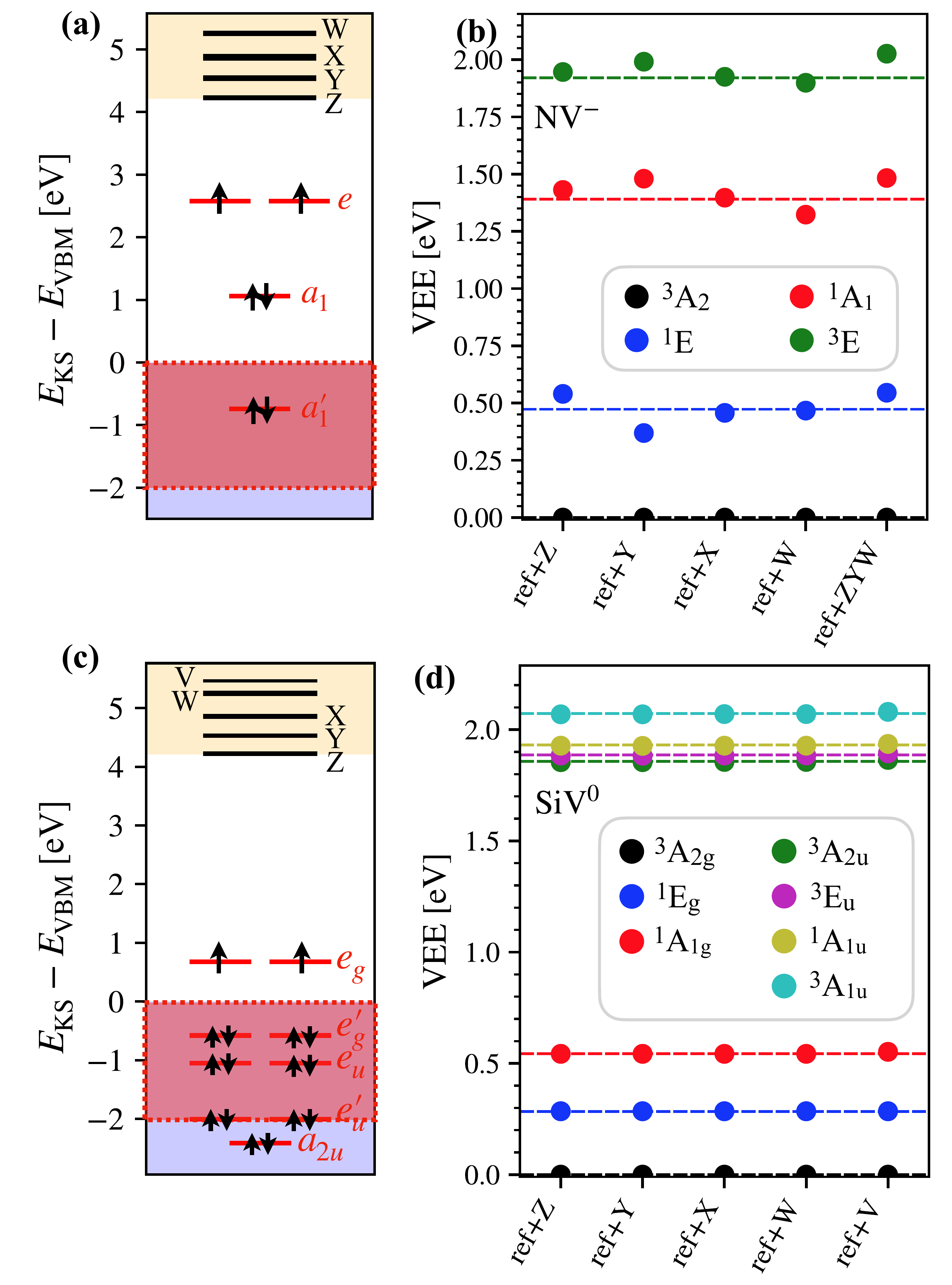}
\caption{\label{fig:unocc_effect} Effect on computed vertical excitation energies (VEEs) of adding unoccupied orbitals into the active space, in 511-atom supercell calculations for the NV$^-$(upper panels) and SiV$^0$ (lower panels) in diamond. 
(a) and (c) illustrate the chosen reference active space by red lines and red shades; it also shows groups of unoccupied orbitals added to the reference active space by letters. $E_\mathrm{KS}$ and $E_\mathrm{VBM}$ denote Kohn-Sham (KS) eigenvalues and the valence band maximum (VBM), respectively.
(b) and (d): Comparison of VEEs computed with different active spaces, which are constructed by adding 1--3 groups of unoccupied KS orbitals to a reference active space without unoccupied orbitals.
Reference active space (``ref'') in both defects is the minimum model plus occupied orbitals with KS energies above VBM\CHANGES{$-2$} eV.
VEEs of the reference active space are marked with dashed horizontal lines.
The \CHANGES{MR-CISD} impurity solver (see text) is used throughout.
}
\end{figure}

We begin by studying the effects of including unoccupied orbitals for the NV$^-$ center in a 511-atom supercell. The reference active space includes the 4 minimum model orbitals and all occupied orbitals with a KS energy of at least \CHANGES{2} eV below the VBM.
We then select four groups of KS orbitals above the Fermi energy, with each group being at least 0.1 eV apart from each other, as shown in Fig.~\ref{fig:unocc_effect}(a). These groups of KS orbitals are named Z, Y, X, and W, from lowest to highest KS energy, respectively, and they each contain six orbitals. We add each group\CHANGES{, one at a time,} to the reference active space, perform a QDET calculation, then diagonalize the effective Hamiltonian with \CHANGES{an MR-CISD} solver. The VEE results are given in Fig.~\ref{fig:unocc_effect}(a). Additionally, we \CHANGES{perform calculations where the groups Z, Y, and W are added to the reference active space together, all of which are found to have significant ($>0.04$ eV) effects when added individually.} The effect of unoccupied orbitals is sizeable for the NV center, reaching $\sim$0.1 eV when orbital groups Z, Y, \CHANGES{W} are all added to the active space. 

Next, we study the effects of including unoccupied orbitals in the active space for neutral group IV vacancies in diamond, and we present results for 511-atom supercells. The reference active space is selected in the same way as for the NV center; we then choose 25 KS orbitals above the Fermi level, which we partition into 5 orbital groups, separated in energy from each other by at least 0.1 eV, \CHANGES{and add those groups individually to the reference active space}. Our results are presented in Fig.~\ref{fig:unocc_effect}(c) for SiV$^0$. We find similar trends for all neutral group IV vacancies (see Section S5 in SI for details).

Unlike the case of the NV$^-$ center, the effect of adding unoccupied orbitals is found to be negligible (\CHANGES{$<0.01$} eV) for all orbital groups.

\begin{figure}
\includegraphics[width=1.0\linewidth]{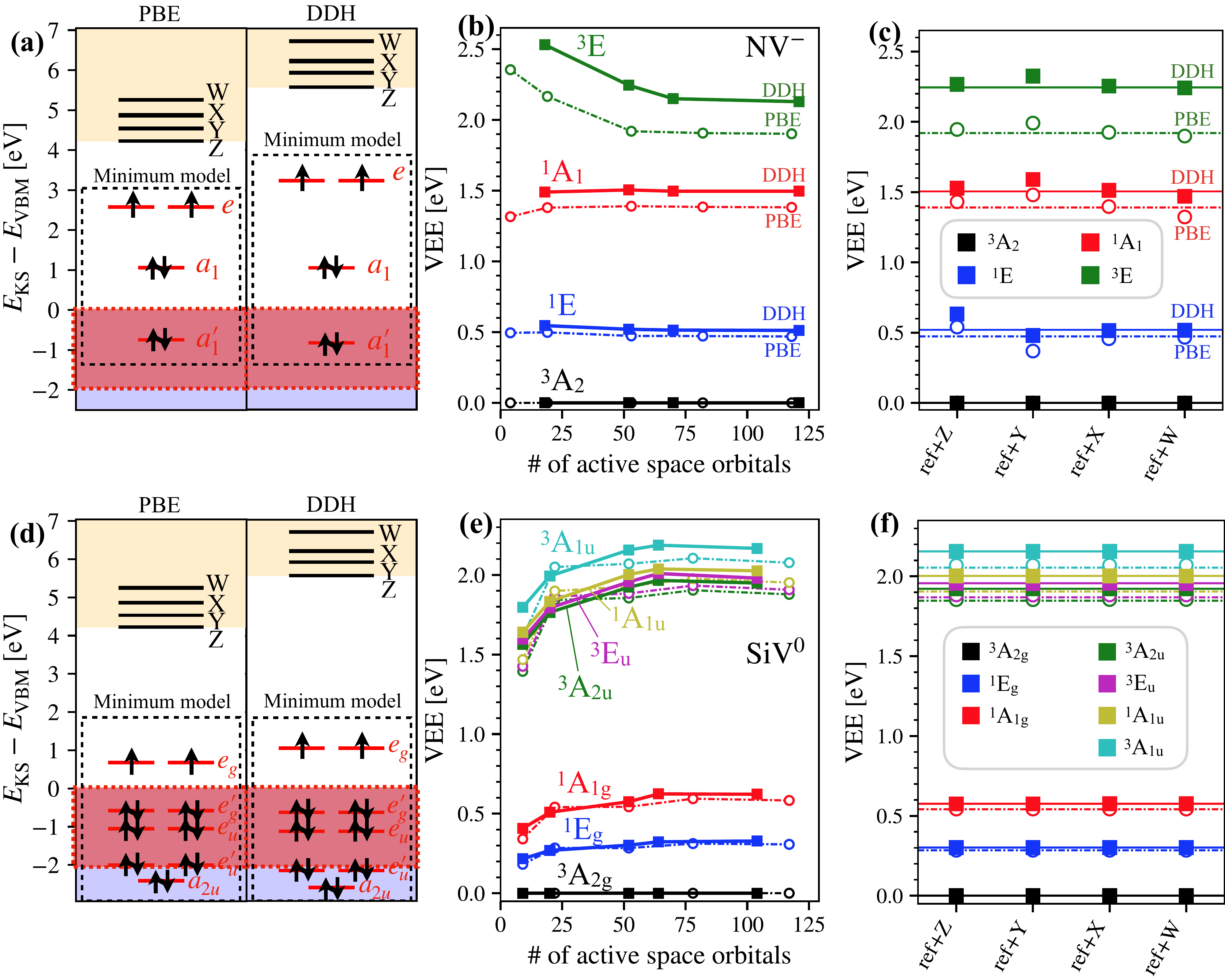}
\caption{\label{fig:newDC_DDH_NV_SiV} (a) Minimum model (dashed-line black box), reference active space (red lines and shades), and unoccupied orbital groups (solid black lines marked with $\mathrm{W,X,Y,Z}$) in the NV$^-$ center in diamond, using the PBE and DDH functionals; (b) Comparison of computed many-body states using the DC2025 double-counting scheme (see text) and the PBE functional (dash-dotted lines) and DDH functional (solid lines), where the vertical excitation energies (VEEs) are reported as a function of the number of orbitals in the active space; 
(c) Comparison of VEEs computed with the PBE and DDH functional and with different active spaces, which are constructed by adding each group of unoccupied Kohn-Sham (KS) orbitals to a reference active space of occupied orbitals. The PBE results are shown as empty circles and the DDH results are shown as filled squares.
Reference active space (``ref'') is the minimum model plus occupied orbitals with KS energies up to the VBM$-2$ eV for each functional.
VEEs of the reference active space are marked with dash-dotted horizontal lines (PBE) or solid horizontal lines (DDH).
(d)--(f) show  the same results as (a)--(c), respectively, but for the SiV$^0$ defect.
The MR-CISD impurity solver (see text) is used throughout.
}
\end{figure}

\CHANGES{
\subsection{Use of hybrid functionals in DFT calculations}\label{subsec:ddh}

To assess the dependence of QDET results on the choice of the exchange-correlation functional used in DFT calculations, we carried out calculations for the NV$^-$ and SiV$^0$ in diamond using the DDH hybrid functional \cite{Skone_2014_PRB,Skone_2016_PRB}, which incorporates a fraction of exact exchange with the mixing parameter $\alpha = 0.18$ equal to the inverse of the macroscopic dielectric constant of bulk diamond. The calculations used the adaptively compressed exchange operator, a low-rank representation of the exact exchange, to reduce the computational cost\cite{Lin_2016_JCTC,Yu_2025_JCP}.
The convergence of the VEEs as a function of the number of orbitals in the active space and the effect of including unoccupied KS orbitals are reported in Fig.~\ref{fig:newDC_DDH_NV_SiV}.
For comparison, PBE results are also shown. We find that for NV$^-$, the VEEs obtained with the DDH functional are generally higher than the corresponding PBE results, especially for the ${}^3\!E$ state, where the difference is $\sim$ 0.2 eV. The effect of each unoccupied orbital group in the VEE of the NV$^-$ is similar when computed with PBE and DDH. For SiV$^0$, the VEEs computed with the PBE and DDH functionals are similar (maximum difference $<0.1$ eV), and for both functionals, the effects of including unoccupied orbitals in the active space are negligible ($<0.01$ eV).
}

\subsection{Effect of hybridization}

\begin{figure}
\includegraphics[width=0.8\linewidth]{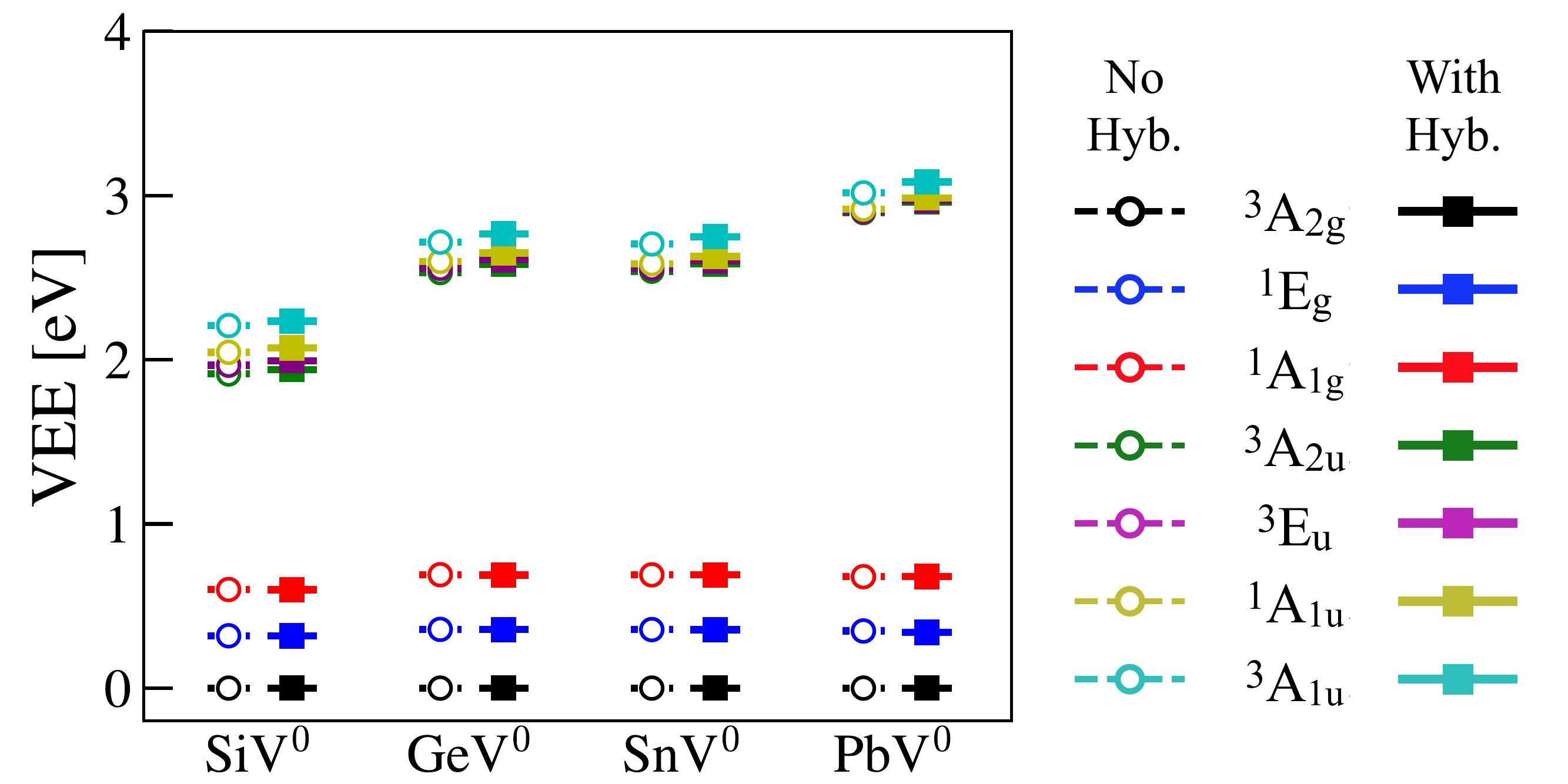}
\caption{\label{fig:newDChyb_groupIV}Effect of adding hybridization terms in the calculation of vertical excitation energies (VEEs) of neutral group IV vacancies in diamond. The VEE values are obtained in a cell with 215 atoms, without and with hybridization, and in the largest active space size we computed for each defect (see Fig. S5 in SI for the full convergence plot).
DFT calculations were carried out with the PBE functional, and DC2025 was used throughout. 
We used FCI and MR-CIS(D) as impurity solvers in the absence and presence of hybridization, respectively.}
\end{figure}

Here we compare results obtained with and without hybridization, and for these comparisons, we\CHANGES{ use a 215-atom supercell and} do not include unoccupied orbitals in the active space. We begin by discussing the neutral group IV vacancies.
As described in the theory section, we select bath orbitals starting from those with the largest contributions $\mathcal{S}_b$ to the hybridization function $\Delta(\omega)$, until $\sum_b \mathcal{S}_b$
reaches the threshold $\mathcal{T}=2/3$ (66.67\%), which we verified is sufficient to obtain converged results. (See Section S4 in SI for convergence tests.)
We then adjust the number of bath orbitals, if necessary, to ensure that the number of electrons in the active space is correct. 
Fig.~\ref{fig:newDChyb_groupIV} shows our results for the largest active space size. The VEEs with approximate hybridization show only small differences compared to the case without hybridization, with no change in state ordering, and 
the largest effect (in PbV$^0$) is about +0.07 eV.

The case of the NV$^-$ center is more complex.
For all choices of active space studied here, we added four extra electrons to the number of electrons initially selected (Section~\ref{subsec:bath_orb}), to ensure that the number of electrons in the active space is correct. For each given excited state we are interested in, we obtain multiple solutions of the auxiliary Hamiltonian (defined in the active plus bath orbital space) that could all have projection onto this given state in the active space.
(see Section S6 in SI).
In other words, in the combined Hilbert space $A\oplus B$, multiple states (e.g. $|a\rangle \otimes|b_1\rangle$, $|a\rangle \otimes|b_2\rangle$; $|b_1\rangle$ and $|b_2\rangle$ are in $B$) correspond to a single state in the original Hilbert space $A$ ($|a\rangle$). 
In addition, states of the form $|a\rangle\otimes|b\rangle$ can mix with states whose projection in the active space is different from the state $|a\rangle$.

\begin{figure}
\includegraphics[width=0.8\linewidth]{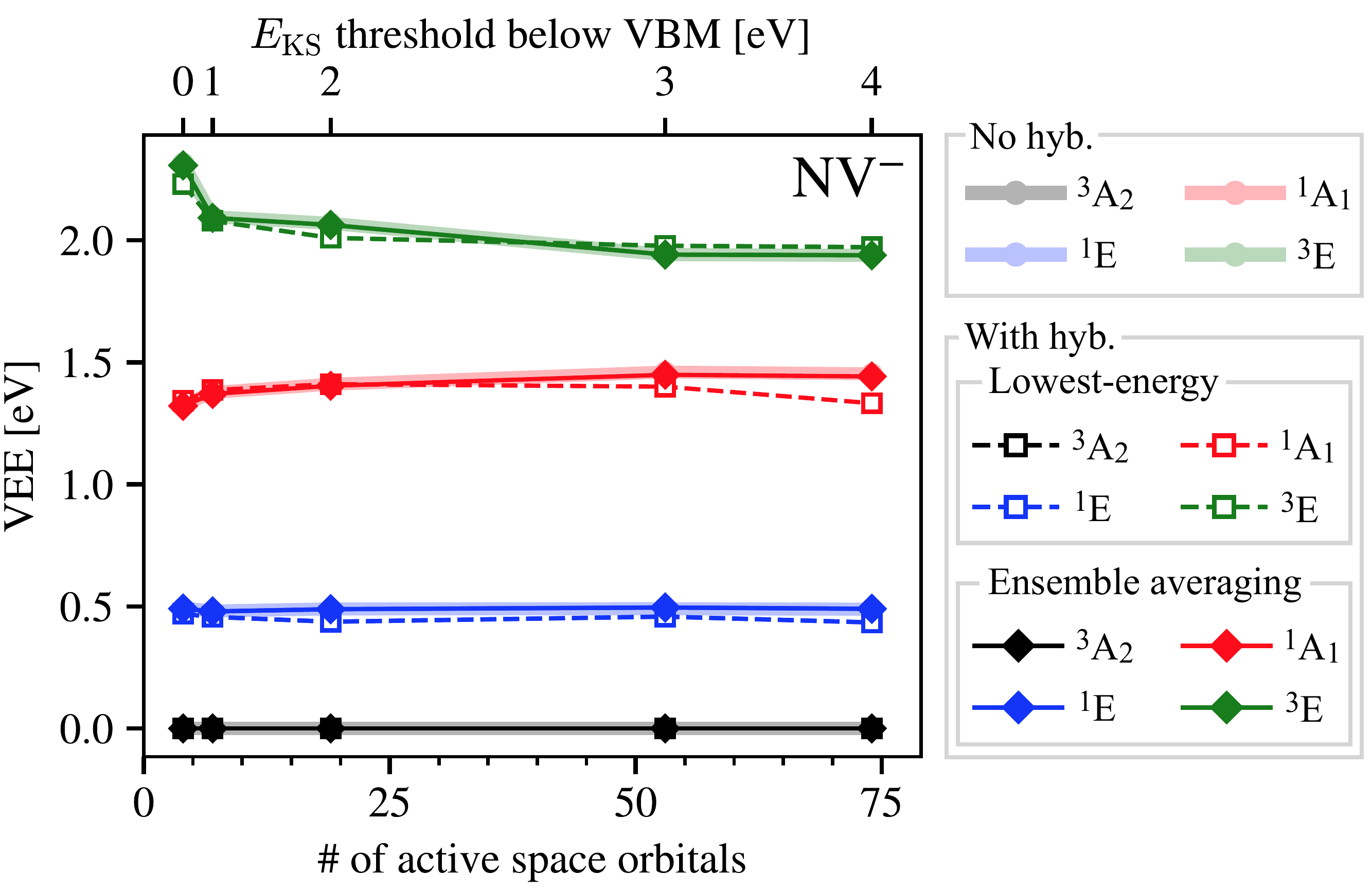}
\caption{\label{fig:newDChyb_NV}Comparison of the vertical excitation energy computed without hybridization (thick shades) and with hybridization (solid/dashed lines), as a function of the number of orbitals in the active space\CHANGES{ and as a function of the Kohn-Sham energy ($E_\mathrm{KS}$) threshold below the VBM}, for the NV$^-$ in diamond, computed with a 215-atom supercell. DFT calculations are carried out with the PBE functional, and we used the refined double counting scheme (DC2025, see text). We used FCI and CIS(D) as impurity solvers in the absence and presence of hybridization, respectively. Due to the interaction between the active space and the bath introduced by the hybridization, an excited state with a given symmetry may split into multiple states. Dashed and solid lines show two approaches used to take into account multiple states (see text). Dashed lines show the lowest-energy state. Solid lines show the ensemble-averaged VEE (Eq.~\ref{eq:ensemble_average}).}
\end{figure}

One may then choose the lowest energy state wherever there is a splitting of states. This choice is consistent with our approach without hybridization, and corresponds to the ``zero-temperature'' limit.
We see from Fig.~\ref{fig:newDChyb_NV} that the hybridization effects are small but not negligible (e.g. $\sim0.1$ eV for state ${}^1A_1$).

Another way to resolve the issue of split states is to consider that we are measuring an ensemble of states using the target $i$-th excited-state $|\tilde P_i\rangle$ as a basis, with the probability of each state in the ensemble proportional to the squared overlap with $|\tilde P_i\rangle$. Here, $|\tilde P_i\rangle$ is constructed as $|P_i\rangle \bigotimes |1\ldots 10\ldots 0\rangle$ for spins up and down, and is a direct product of the non-hybridized excited-state $|P_i\rangle$ with a bath state, obtained by filling bath orbitals from the lowest energy upwards. Such a state mimics the expected excited state in the full system, and the formula for the ``ensemble-averaged'' energy is given by

\begin{equation}
\label{eq:ensemble_average}
E^\mathrm{ens}_i = \frac{\sum_j |\langle P_i|\Psi_j\rangle|^2 E_j}{\sum_j |\langle P_i|\Psi_j\rangle|^2}.
\end{equation}

The VEEs obtained by this ensemble-average formalism are shown as solid lines in Fig.~\ref{fig:newDChyb_NV}. The hybridization effect is negligible after we perform the ensemble average, at most 12 meV.

\CHANGES{

Having examined the effects of double-counting, unoccupied orbitals, exchange-correlation functional used in DFT calculations, and hybridization, we now present the computed VEEs of NV$^-$ obtained by incorporating all these contributions in Table~\ref{table:NV_best_results}.

Our DDH results are further compared with those obtained using other embedding methods~\cite{Haldar_2023_JPCL,Bockstedte_2018_npjQM,Martirez_2025_arXiv} in Table~\ref{table:NV_literature}. We find that using the DDH functional instead of PBE and including unoccupied orbitals improve the agreement with experiment. Our final results not only agree qualitatively with those of prior studies, but also show good agreement with experimental data.
}

\begin{table}[h!]
\centering
\begin{tabular}{|c|c|c|c|c|c|}
\hline
State &
\multicolumn{2}{c|}{$\mathrm{VEE}_\mathrm{occ}$ [eV]} &
$\Delta$VEE$_\mathrm{unocc}$[eV] &
\multicolumn{2}{c|}{VEE [eV]} \\
\cline{2-3}\cline{5-6}
& PBE & DDH & & PBE & DDH \\
\hline
${}^1\!E$ & 0.469 (0.473) & 0.512 (0.520) & +0.072 & 0.541 & 0.584 \\
${}^1\!A_1$ & 1.382 (1.390) & 1.496 (1.505) & +0.093& 1.475 & 1.589 \\
${}^3\!E$ & 1.902 (1.920) & 2.129 (2.245) & +0.106 & 2.008 & 2.235 \\
\hline
\end{tabular}
\caption{Vertical excitation energies (VEEs) computed for the NV$^-$ center in diamond using a 511-atom supercell. The values of $\mathrm{VEE}_\mathrm{occ}$ in the second and third columns are obtained with the active space MM+VBM$-$3eV (minimum model [MM] plus orbitals with KS energy up to 3 eV below the valence band maximum [VBM]), using either PBE or DDH in DFT calculations. Within parentheses, we give results obtained with an active space MM+VBM$-$2eV (MM plus orbitals with KS energy up to 2 eV below VBM). The value $\Delta \mathrm{VEE}_\mathrm{unocc}$ is the difference in energy between VEE$_\mathrm{occ}$ at the PBE level computed with the MM+VBM$-$2eV active space and by adding to the occupied active space orbitals 18 unoccupied orbitals (groups ZYW in Fig.~\ref{fig:unocc_effect}). The effect of hybridization was omitted, as it amounts to at most 12 meV. The VEEs at the PBE and DDH level given in the 5th and 6th columns are computed as $\mathrm{VEE}_\mathrm{occ} + \Delta \mathrm{VEE}_\mathrm{unocc}$.}
\label{table:NV_best_results}
\end{table}

\begin{table}[h!]
\centering
\begin{tabular}{ |c|c|c|c|c|c| } 
\hline
State & This work & NEVPT2-DMET\cite{Haldar_2023_JPCL} & CI-CRPA\cite{Bockstedte_2018_npjQM} & emb-MRPT2\cite{Martirez_2025_arXiv} & Experiment\cite{Davies_1976_PRSA,Rogers_2008_NJP,Goldman_2015_PRL,Kehayias_2013_PRB,Goldman_2015_PRB} \\
& [eV] & [eV] & [eV] & [eV] & [eV] \\
\hline
${}^1\!E$ & 0.584 & 0.50 & 0.49 & 0.58--0.71 & 0.34--0.43 (ZPL) \\
${}^1\!A_1$ & 1.589 & 1.56 & 1.41 & 1.87--1.98 & 1.51--1.60 (ZPL) \\
${}^3\!E$ & 2.235 & 2.31 & 2.02 & 2.23--2.30 & 1.945 (ZPL), 2.18 (VEE) \\
\hline
\end{tabular}
\caption{Comparison of vertical excitation energies (VEEs) of the NV$^-$ center in diamond obtained in this work [VEE (DDH) from Table~\ref{table:NV_best_results}] and published results. The calculation of density matrix embedding theory with n-electron valence state second-order perturbation theory impurity solver (NEVPT2-DMET) used a 215-atom supercell and DFT calculations were carried out with PBESol.\cite{Haldar_2023_JPCL} The calculation with the constrained random-phase approximation and a configuration interaction impurity solver (CI-CRPA) used a 511-atom supercell and carried out DFT calculations with the HSE06 functional.\cite{Bockstedte_2018_npjQM} The calculation of embedded multi-reference second-order perturbation (emb-MRPT2) used a 63-atom supercell and DFT calculations with the HSE06 or r$^2$SCANL functionals.\cite{Martirez_2025_arXiv} The zero-phonon line (ZPL) was experimentally determined to be at 1.945 eV through low-temperature absorption and emission experiments.\cite{Davies_1976_PRSA} Adding the computed excited-state Franck-Condon shift obtained with $GW$-BSE\cite{Ma_2010_PRB} to the experimental ZPL yields a VEE of 2.18 eV, consistent with the highest-intensity peak of the broad phonon side band in the absorption spectrum measured at liquid-nitrogen temperature.\cite{Davies_1976_PRSA}}
\label{table:NV_literature}
\end{table}

\subsection{Comparison between different impurity solvers}

Here, we distinguish two cases: (i) adding unoccupied states to the active space and (ii) including a hybridization term, when computing VEEs.

For case (i), we compare the results obtained with four impurity solvers for the NV center in a 215-atom supercell. We considered two different active spaces, and we report our results in Table~\ref{table:imp_solver_unocc}. Comparing the FCI solutions with those of selected CI and MR-CISD, we see that the latter methods are remarkably accurate. The accuracy of AFQMC is slightly worse, but \CHANGES{since the precision of AFQMC is limited to $\simeq$ 0.01 eV (see Section~S7 in SI), the} discrepancy is inconsequential, and the results are all within chemical accuracy (0.04 eV).

\begin{table}[h!]
\centering
\begin{tabular}{ |c|c|c|c|c| } 
\hline
Active space & Solver & $^1\!E$ [eV] & $^1\!A_1$ [eV] & $^3\!E$ [eV] \\
\hline
MM+VBM-1eV+6u & FCI & 0.3833 & 1.4880 & 2.1817 \\ 
 & Selected CI & 0.3844 & 1.4876 & 2.1819 \\ 
 & MR-CISD & 0.3842 & 1.4883 & 2.1820 \\ 
 & AFQMC & 0.39 & 1.50 & 2.14 \\
\hline
MM+VBM-2eV+6u & Selected CI & 0.3822 & 1.5168 & 2.1366 \\
 & MR-CISD & 0.3824 & 1.5182 & 2.1375 \\
 & AFQMC & 0.39 & 1.53 & 2.10
 \\
\hline
\end{tabular}
\caption{Comparison of results for the vertical excitation energies (VEEs) of the NV in diamond, obtained with different impurity solvers. We show results for two different active spaces consisting of (1) minimum model, plus all occupied orbitals with a KS energy higher than VBM-1 eV, plus 6 lowest unoccupied KS orbitals (``MM+VBM-1eV+6u''); (2) same as (1), but for the occupied orbitals, we use those with KS energy higher than VBM-2 eV (``MM+VBM-2eV+6u''). \CHANGES{Precision of AFQMC is limited to 2 digits after the decimal.}
}
\label{table:imp_solver_unocc}
\end{table}

In Table~\ref{table:imp_solver_hyb}, we compare the computed VEEs for an NV center and a SiV$^0$ computed in a 215-atom cell, where in both cases we chose an active space including the minimum model plus all occupied orbitals with KS energy 2 eV below the VBM. Bath orbitals with the largest contribution to $\Delta(\omega)$ are included until the sum of contributions reaches 2/3. 
Here, for computational convenience in the case of a state splitting for the NV, we choose the lowest-energy state for comparison.

We find that the hybridization effect is negligible when using either solver, and their difference is within chemical accuracy.

\begin{table}[h!]
\centering
\begin{tabular}{ |c|c|c|c|c|c|c| } 
\hline
System & $N_A$ & $N_B$ & State & No hyb. [eV] & MR-CIS(D) [eV] & AFQMC [eV] \\
\hline
NV$^-$ & 19 & 41 & ${}^1\!E$ & 0.490 & 0.478 & 0.48 %
\\
 (215-atom) & & & ${}^1\!A_1$ & 1.411 & 1.411 & 1.39 %
\\
 & & & ${}^3\!E$ & 2.070 & 2.028 & 2.05 %
\\
\hline
SiV$^0$ & 20 & 30 & ${}^1\!E_g$ & 0.311 & 0.311 & 0.31 %
\\
 (215-atom) & & & ${}^1\!A_{1g}$ & 0.595 & 0.595 & 0.59 %
\\
 & & & ${}^3\!A_{2u}$ & 2.101 & 2.109 & 2.09 %
\\
 & & & ${}^3\!E_{u}$ & 2.120 & 2.128 & 2.13 %
\\
 & & & ${}^1\!A_{1u}$ & 2.154 & 2.162 & 2.16 %
\\
 & & & ${}^3\!A_{1u}$ & 2.289 & 2.297 & 2.29 %
\\
\hline
\end{tabular}
\caption{Comparison of the MR-CIS(D) and AFQMC impurity solvers in obtaining vertical excitation energies (VEEs) in the presence of hybridization, for the NV$^-$ and SiV$^0$ in diamond. VEE results without hybridization are also listed for comparison. The choice of the active space is specified through the number of active space and bath orbitals ($N_A,N_B$). \CHANGES{Precision of AFQMC is limited to 2 digits after the decimal.} }
\label{table:imp_solver_hyb}
\end{table}

\subsection{Cr(\textit{o}-tolyl)$_4$ molecular qubit}

We finally turn to the application of QDET to the study of a molecular qubit, the Cr(\textit{o}-tolyl)$_4$ \cite{Bayliss_2020_Science,Laorenza_2021_JACS}. 
We use a tetragonal unit cell with lattice constants 11.92$\times$11.92$\times$7.89 Angstrom, which contains one Cr(\textit{o}-tolyl)$_4$ and one Sn(\textit{o}-tolyl)$_4$ molecule.

Table~\ref{table:Cr_o_tolyl} shows the VEE from the ground to the first singlet excited state computed with QDET, using DC2025 and with several different choices of the active space, starting with the two partially occupied KS orbitals as the minimum model and adding KS states close to the band edges. The lowest three unoccupied KS orbitals are the Cr $3d$ orbitals, and we attempt to include them in our active space choices when feasible. 
FCI/selected CI solvers are used to diagonalize the effective Hamiltonian when unoccupied orbitals are absent/present in the active space, respectively.
In all cases, the VEEs computed with QDET are in excellent agreement with experiments (within chemical accuracy) when compared to the experimental zero-phonon line (ZPL) of 1025 nm (1.210 eV)\cite{Bayliss_2020_Science}, after including the Franck-Condon shift (FCS). Note that the FCS was computed using spin-flip time-dependent density functional theory (TDDFT) with the PBE functional by relaxing the atomic geometry in the excited state, yielding a value of 0.031 eV.

\begin{table}[h!]
\centering
\begin{tabular}{ |c|c|c| } 
\hline
Active space choice & Active space size & First VEE [eV] \\
\hline
MM+VBM$-$1 eV & (48e,25o) & 1.193 \\
MM+VBM$-$3 eV & (82e,42o) & 1.204 \\
MM+VBM$-$4 eV & (112e,57o) & 1.206 \\
MM+VBM$-$1 eV+3u & (48e,28o) & 1.202 \\
MM+VBM$-$3 eV+3u & (82e,45o) & 1.225 \\
\hline
& Experiment ZPL & 1.210 \\
& Franck-Condon shift & +0.031 \\
& Experiment VEE & 1.241 \\
\hline
\end{tabular}
\caption{Vertical excitation energies (VEEs) of the first singlet state of the Cr(\textit{o}-tolyl)$_4$ molecular qubit, computed with QDET. Results obtained with different active space choices are compared with experiment\cite{Bayliss_2020_Science}. The experimental VEE is estimated by adding a Franck-Condon shift to the experimental zero-phonon line (ZPL) energy. The Franck-Condon shift is estimated using TDDFT at the DFT-PBE relaxed geometry. We adopted the following notation for active spaces: ``MM'' stands for the minimum model (two partially occupied KS orbitals). ``+VBM$-n$ eV'' denotes that occupied orbitals with Kohn-Sham energy higher than VBM$-n$ eV are added to the MM. ``+3u'' indicates that the three lowest unoccupied KS orbitals, i.e. the unoccupied Cr $3d$ orbitals, are also included in the active space. The second column shows the total number of electrons (for both spins) and the total number of orbitals in the active space.}
\label{table:Cr_o_tolyl}
\end{table}

\section{Conclusion}
In this work, we presented several important improvements to the quantum defect embedding theory. We refined the definition of the double-counting term, including a consistent treatment of the frequency dependence in all terms of the effective Hamiltonian. %
We found differences with our previous results on the order of 0.2 eV, resulting in a better agreement with experiments for the case of the neutral vacancies in diamond. 
We investigated the effect of including unoccupied orbitals in the active space, neglected in our previous work\cite{Sheng_2022_JCTC}, and found a negligible impact on the electronic structure of neutral group IV vacancies. However, the effect on the many-body states of the NV center is non-negligible, being on the order of 0.1 eV and improving the agreement with experiment. 
\CHANGES{Further, our QDET results show a moderate dependence on the exchange-correlation functional used in DFT calculations, with the DDH functional generally yielding higher (up to 0.2 eV) VEEs than the PBE functional. In the case of the NV$^-$ center, the DDH results are in better agreement with experiment than the PBE ones.}
\CHANGES{Finally, }%
we included in QDET an approximate 
hybridization term, through the definition of extra bath orbitals. We found that hybridization effects are within 0.07 eV for group IV defects in diamond, and 12 meV for the NV center. 
It remains to be seen whether the negligible hybridization effects found here become more substantial once self-consistency in the hybridization is included. Work is in progress to add self-consistency to the current QDET framework. 
Other factors, such as a frequency-dependent screened Coulomb potential, could also be contributors to the remaining difference from experiments.
We also verified the consistency of results obtained with different impurity solvers, and we extended the application of QDET to molecular qubits.
In summary, our work %
represents an important step in the development of embedding methods 
applicable to large supercells representing defects in solids.

\begin{acknowledgement}

The authors thank Diego Sorbelli, Chia-Nan Yeh, Shiwei Zhang, and Huanchen Zhai for useful discussions.
This work was supported by the Midwest Integrated Center for Computational Materials (MICCoM). MICCoM is part of the Computational Materials Sciences Program funded by the U.S. Department of Energy, Office of Science, Basic Energy Sciences, Materials Sciences, and Engineering Division through the Argonne National Laboratory, under contract No. DE-AC02-06CH11357.
This research used resources of the National Energy Research Scientific Computing Center (NERSC), a DOE Office of Science User Facility supported by the Office of Science of the U.S. Department of Energy under contract No. DE-AC02-05CH11231 using NERSC award ALCC-ERCAP0025950, and resources of the University of Chicago Research Computing Center.

\end{acknowledgement}

\begin{suppinfo}

The Supporting Information is available free of charge at ...

Details of theoretical derivation (section 1), convergence versus PDEP (section 2), comparison of active space selection criteria (section 3), cutting off bath orbitals (section 4), convergence as a function of the active space size in neutral group IV vacancies in diamond (section 5), state splitting due to hybridization in NV$^-$@Diamond (section 6), comment on excited state AFQMC calculation (section 7)

\end{suppinfo}
\bibliography{main}

\end{document}


Throughout this Supplementary Information, the notations defined in Section 2
of the main text are still applicable.

\section{Details of theoretical derivation}
\subsection{Derivation for $W_A(\omega)$}

We prove a mathematical relation which is used in our derivations in the main text.
Our starting point is the equation that relates the inverse of a block matrix to the original block matrix:

\begin{align}
    \begin{bmatrix}
        \mathcal{A} & \mathcal{B} \\
        \mathcal{C} & \mathcal{D}
    \end{bmatrix}^{-1}
    &=
    \begin{bmatrix}
    \mathcal{J}^{-1} & -\mathcal{J}^{-1} \mathcal{B} \mathcal{D}^{-1} \\
    -\mathcal{D}^{-1} \mathcal{C} \mathcal{J}^{-1} & \mathcal{D}^{-1}+\mathcal{D}^{-1} \mathcal{C} \mathcal{J}^{-1} \mathcal{B} \mathcal{D}^{-1}
\end{bmatrix}\label{eq:matinv1}\\
    &=
    \begin{bmatrix}
    \mathcal{A}^{-1}+\mathcal{A}^{-1} \mathcal{B}
    \mathcal{K}^{-1}\mathcal{C}\mathcal{A}^{-1} & -\mathcal{A}^{-1} \mathcal{B} \mathcal{K}^{-1} \\
    -\mathcal{K}^{-1} \mathcal{C} \mathcal{A}^{-1} & \mathcal{K}^{-1}
    \end{bmatrix},\label{eq:matinv2}
\end{align}
where
$\mathcal{J}=\mathcal{A}-\mathcal{B} \mathcal{D}^{-1} \mathcal{C}$ and $\mathcal{K}=\mathcal{D}-\mathcal{C} \mathcal{A}^{-1} \mathcal{B}$ is the Schur complement of $\mathcal{D}$ and $\mathcal{A}$, respectively.
The first and second equations hold if $\mathcal{J}$ and $\mathcal{K}$ are invertible, respectively.
We will assume that this is always the case for a matrix defined in the $A\oplus E$ space.

Applying Eqs.~\ref{eq:matinv1} and \ref{eq:matinv2}, it is straightforward to prove the following identity,

\begin{equation}
[(M_A)^{-1}-N_A]^{-1} = \left\{\left[M^{-1}-N\right]^{-1}\right\}_A\,,\label{eq:inv_identity}
\end{equation}
where $M, N$ are both matrices of dimension $A\oplus E$, but $N$ is only defined (nonzero) in the $A$ subspace.

The identity in Eq.~\ref{eq:inv_identity} is equivalent to the following statement,
\begin{equation}
F^{-1} = M^{-1} - N\,\,\,\, \iff\,\,\,\,
[F_A]^{-1} = [M_A]^{-1} - N_A\,,\label{eq:inv_statement}
\end{equation}
for matrices $F,M,N$ of dimension $A\oplus E$, if $N$ is only defined (nonzero) in the $A$ subspace.

Applying Eq.~\ref{eq:inv_statement}, it is therefore straightforward to show that

\begin{equation}
  [W_{A}(\omega)]^{-1} = \left[ W^R_{A}(\omega) \right]^{-1} - P^{\mathrm{HIGH}}_A(\omega),
\end{equation}
since $W^{-1}(\omega) = \left[ W^R(\omega) \right]^{-1} - P^{\mathrm{HIGH}}(\omega)$,
and $P^\mathrm{HIGH}(\omega)$ is only defined in $A$.

\subsection{Derivation for $W^\mathrm{eff}(\omega)$}

In Section 2.2 of the main text, we write the double-counted self-energy as

\begin{equation}
  \Sigma^\mathrm{DC}_A(\omega) =v^\mathrm{eff}\rho_{0,A} + \mathrm{i}\int\mathrm{d}\omega' G_{0,A}(\omega+\omega') W^\mathrm{eff}_A(\omega')\,.\label{eq:dc_sigma}
 \end{equation}

To compute $W^\mathrm{eff}_A(\omega)$, substitute $v$ with $v^\mathrm{eff}$ and $P_0$ with $P_{0,A}$ in the expression $W_0(\omega) = [v^{-1} - P_0(\omega)]^{-1}$, and note that $v^\mathrm{eff}=W^R_A(\omega=0)$ in second quantization. This leads to

\begin{align}
W^\mathrm{eff}_A(\omega) &= [(v^\mathrm{eff})^{-1} - P_{0,A}(\omega)]^{-1}\nonumber\\
&= \left\{[v^{-1} - P_0(\omega=0) + P^\mathrm{DC}(\omega=0) - P_0^A(\omega)]^{-1}\right\}_{A}\nonumber\\
&= \left\{[v^{-1} - P_0^R(\omega=0) - P_0^A(\omega)]^{-1}\right\}_{A}.\label{eq:dc_W}
\end{align}
where we applied Eq.~\ref{eq:inv_identity}.

Here, the inversion of a four-index quantity (for example $v$) proceeds as follows: first, each pair of indices corresponding to the same vertex on the Feynman diagram is packed into a single generalized index (as is a common practice in Bethe-Salpeter equations (BSE) calculations). This transforms the original four-index quantity into a (two-index) matrix, which can then be inverted using standard matrix inversion techniques. Finally, a reverse procedure (unpacking) is performed to recover the four-index quantity.
For example, suppose that we want to compute $\mathfrak{V}=v^{-1}$. We first convert the tensor $v_{ijkl}$ into a matrix $\tilde v_{pq}$ where $p$ packs the indices $\{i,k\}$ and $q$ packs the indices $\{j,l\}$. If $i,j,k,l$ are of dimension $N$, the $p,q$ indices will be of dimension $N^2$.  We then invert this $\tilde v$ matrix: $\mathfrak{\tilde V} = {\tilde v}^{-1}$. The tensor $\mathfrak{V}_{ijkl}$ can then be recovered from the matrix $\mathfrak{\tilde V}_{pq}$ by reversing the packing procedure: $p$ is unpacked into $i$ and $k$, and $q$ is unpacked into $j$ and $l$.

Since only $W^\mathrm{eff}_A(\omega)$ is used to derive the double-counting relations, in principle we do not need to define
$W^\mathrm{eff}(\omega)$ outside $A$. However,
for convenience and for comparison with Ref.~\citenum{Sheng_2022_JCTC}, we still choose to \textit{formally} define $W^\mathrm{eff}(\omega)$ in $A\oplus E$ space,
\begin{equation}
W^\mathrm{eff}(\omega) = [v^{-1} - P_0^R(\omega=0) - P_0^A(\omega)]^{-1}\,,
\end{equation}
which yields the same $W^\mathrm{eff}_A(\omega)$ as in Eq.~\ref{eq:dc_W}.

\section{Convergence as a function of the number of PDEP functions}

\begin{figure}
\includegraphics[width=0.7\linewidth]{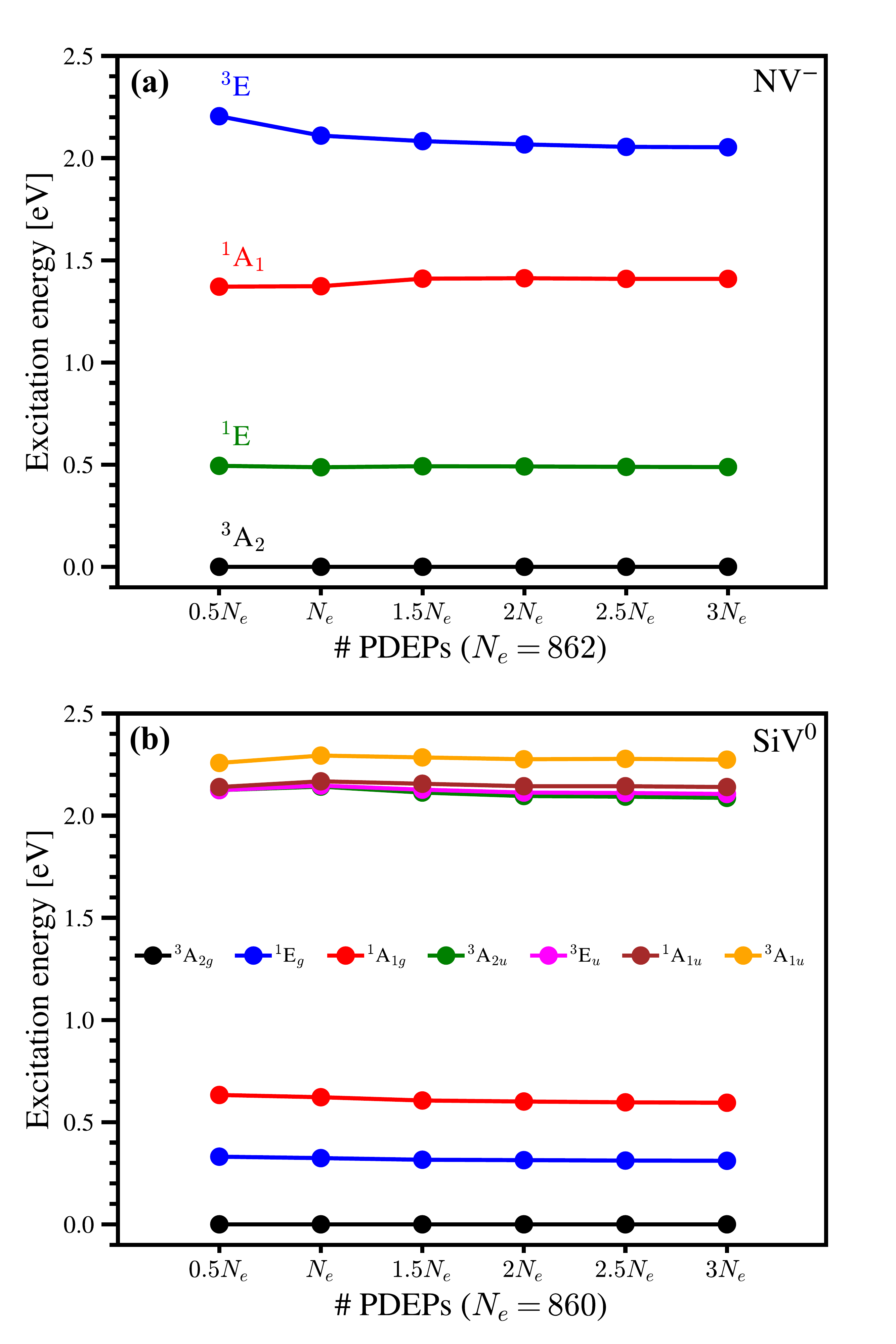}
\caption{\label{fig:supp_pdep_convergence}Vertical excitation energies in a 215-atom supercell (a) $\mathrm{NV^-}$ and (b) $\mathrm{SiV^0}$ in diamond, as a function of the number of projective dielectric eigenpotentials (PDEPs) used in the calculation of dielectric screening. $N_e$ is the total number of electrons.}
\end{figure}

In Fig.~\ref{fig:supp_pdep_convergence}, we investigate the number of projective dielectric eigenpotentials (PDEPs), $N_\mathrm{PDEP}$, necessary to converge the vertical excitation energies, in a 215-atom supercell for $\mathrm{NV^-}$ and $\mathrm{SiV^0}$ in diamond. The active space is the minimal model plus all occupied bands above VBM$-2$ eV, for both defect systems. Convergence within 10 meV is achieved when the number of PDEPs reaches 2--2.5 times the number of electrons in both systems. To account for possible fluctuations in other systems, all calculations reported in the main text use $N_\mathrm{PDEP}=3N_e$ throughout.

\section{Comparison of active space selection criteria}

\begin{figure}
\includegraphics[width=1\linewidth]{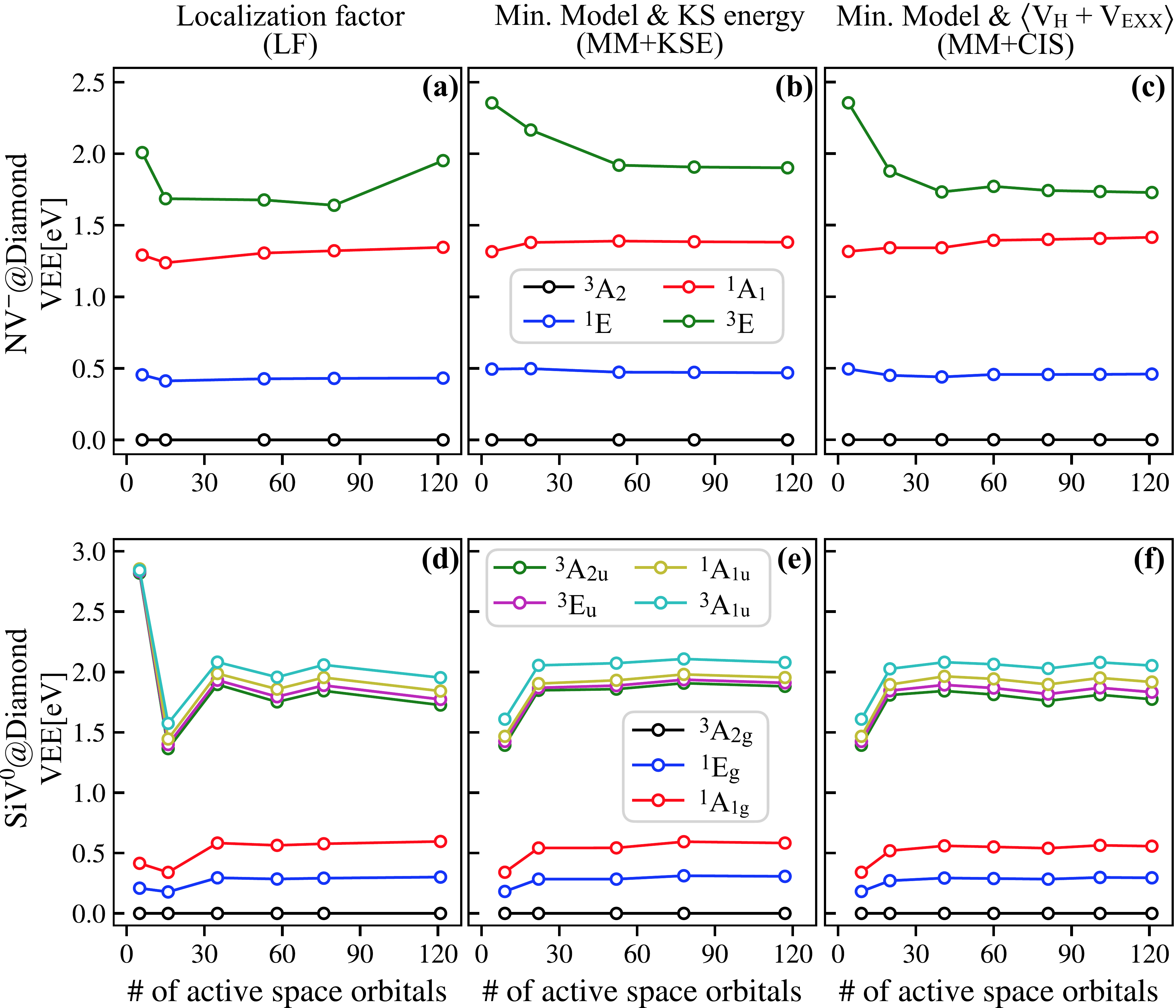}
\caption{\label{fig:supp_compare_crit}Comparison of criteria used to select the active space; in (a)(d) the localization factor is used for the NV$^-$ and SiV$^0$ in diamond, respectively; in (b)(e) the MM+KSE criterion is used for the two defects, respectively; and in (c)(f) the MM+CIS criterion is used for the two defects, respectively. Calculations were carried out in a 511-atom supercell. Vertical excitation energies (VEE) of multi-reference states are plotted against the number of orbitals in the active space.}
\end{figure}

\CHANGES{
In addition to the two active space selection criteria suggested in the main text, we consider another criterion, which we denote as minimum model plus Coulomb interaction strength (MM+CIS), that identifies the orbitals belonging to the active space as follows:

\begin{itemize}
\item The minimum model orbitals (same as those chosen in the MM+KSE criterion in the main text) are included,
\item Orbitals $\zeta_i$ with the highest $\lambda_i$, where $\lambda_i = \sqrt{\sum_{j\in \mathrm{MM}}(\langle \zeta_j| V^H + V^\mathrm{EXX} |\zeta_i\rangle)^2}$, in which the Hartree potential $V^H$ and exact exchange potential $V^\mathrm{EXX}$ are evaluated at the DFT level, are included as well.
\end{itemize}

Here we compare the two active space selection criteria described in the main text (localization factor [LF], and minimum model plus Kohn-Sham energy proximity to VBM [MM+KSE]) and the MM+CIS criterion. In Fig.~\ref{fig:supp_compare_crit}, we report calculations in a 511-atom cell for the NV$^-$ and SiV$^0$ in diamond. We find that both the MM+KSE and MM+CIS criteria lead to smoother convergence curves of VEEs than those obtained with the localization factor criterion, as a function of the size of the active space. Furthermore, for the NV$^-$, the MM+CIS criterion is observed to converge faster than the MM+KSE one, while the convergence rates are similar in the case of SiV$^0$.

However, we found two notable issues when using the MM+CIS criterion in Cr(\textit{o}-tolyl)$_4$ molecular qubit. First, it introduces semi-core orbitals into the active space, which are poorly described by the DFT+$G_0W_0$ method, leading to an abrupt shift of VEEs as a function of the number of orbitals in the active space. Second, the inclusion of the three Cr $3d$ unoccupied orbitals leads to the prediction of a wrong ground state. We notice that if we define the active space as the union of the set of orbitals selected by MM+KSE, and the set of orbitals selected by MM+CIS, the ground state and the VEE of the first singlet are correct. This indicates that the MM+CIS criterion may have missed some important occupied orbitals in the definition of the active space. In conclusion, despite performing well for diamond defects, the MM+CIS criterion might not be applicable in general, hence the more robust MM+KSE criterion is used in our work.
}

\section{Cutting off bath orbitals}
We select bath orbitals as described in Section 3.3 of the main text. We perform the following tests to verify the validity of our selection: for a given active space, we compute all poles $b$ and sort them according to their $\mathcal{S}_b$. Then we perform two tests:

\begin{itemize}
\item Test 1: Use an increasingly large threshold $\mathcal{T}$ to select bath orbitals, then add those to the active space.
\item Test 2: Divide all poles into several groups ($B_1,B_2,\ldots$) with about the same number of poles in each group, then add one group at a time as bath orbitals to the active space ($A\oplus B_1,A\oplus B_2,\ldots$). We allow small changes in the number of poles per group because it is not optimal to break up pole clusters that are almost degenerate in frequency.
\end{itemize}

Table~\ref{table:supp_bath_orb_crit} shows the result of these two tests, for a 215-atom supercell representing $\mathrm{NV^-}$ in diamond. The starting active space consists of 9 bands (7 occupied bands and 2 unoccupied bands closest to the Fermi level). No splitting of states is observed for this active space size. In test 1, we choose the threshold of $\frac{2}{3},\frac{5}{6}$, and $\frac{11}{12}$. In test 2, a total of 461 poles/bath orbitals are first sorted according to their contribution, then divided into 9 groups, each containing 52 bath orbitals (the last one contains 45 orbitals). In test 1, we see most of the hybridization effect is present at the $\frac{2}{3}$ threshold. In test 2, we see the hybridization effect is large for only the first group of poles, and quickly decreases to zero for the latter groups. Both tests justify our criterion and our sum of contribution threshold of 2/3.

\begin{table}[!htbp]
\centering
\begin{tabular}{ |c|c|c|c|c| }
\hline
Threshold & $N_B$ & VEE of ${}^1\!E$ [eV] & VEE of ${}^1\!A_1$ [eV] & VEE of ${}^3\!E$ [eV]\\
\hline
0 & 0 & 0.4780 & 1.3773 & 2.0967 \\
2/3 & 21 & 0.4583 & 1.3896 & 2.0845 \\
5/6 & 53 & 0.4576 & 1.3880 & 2.0868 \\
11/12 & 101 & 0.4577 & 1.3880 & 2.0877 \\
\hline
Group\# & $N_B$ & VEE of ${}^1\!E$ [eV] & VEE of ${}^1\!A_1$ [eV] & VEE of ${}^3\!E$[eV] \\
\hline
1 & 52 & 0.4576 & 1.3880 & 2.0868 \\
2 & 52 & 0.4780 & 1.3773 & 2.0976 \\
3 & 52 & 0.4780 & 1.3776 & 2.0973 \\
4 & 52 & 0.4780 & 1.3770 & 2.0976 \\
5 & 52 & 0.4783 & 1.3774 & 2.0967 \\
6 & 52 & 0.4780 & 1.3773 & 2.0967 \\
7 & 52 & 0.4780 & 1.3773 & 2.0967 \\
8 & 52 & 0.4780 & 1.3773 & 2.0967 \\
9 & 45 & 0.4780 & 1.3773 & 2.0967 \\
\hline
\end{tabular}
\caption{Convergence as a function of the number of bath orbitals, for calculations carried out in a 215-atom supercell for the $\mathrm{NV^-}$ center in diamond, using a 7-occupied-orbital plus 2-unoccupied-orbital active space. In the first column of the upper part of the table, the threshold $\mathcal{T}$ is gradually increased towards one. This increases the number of bath orbitals $N_B$ (second column) and changes the vertical excitation energies (VEE, third column to fifth column). In the lower part of the table, the bath orbitals are sorted according to their $\mathcal{S}_b$ and then divided into 9 groups. Each group of bath orbitals is then individually added to the active space and used to compute vertical excitation energies.}
\label{table:supp_bath_orb_crit}
\end{table}

\section{Convergence as a function of the active space size in neutral group IV vacancies in diamond}
In the main text, we show only selected results on the convergence of our results with respect to the active space size for neutral group IV vacancies. We also show results for unoccupied bands for $\mathrm{NV^-}$ and $\mathrm{SiV^0}$ in diamond only. In this section, we provide an analog of Figs. 2--4 and Fig. 7 for neutral group IV vacancies.

\begin{figure}
\includegraphics[width=0.9\linewidth]{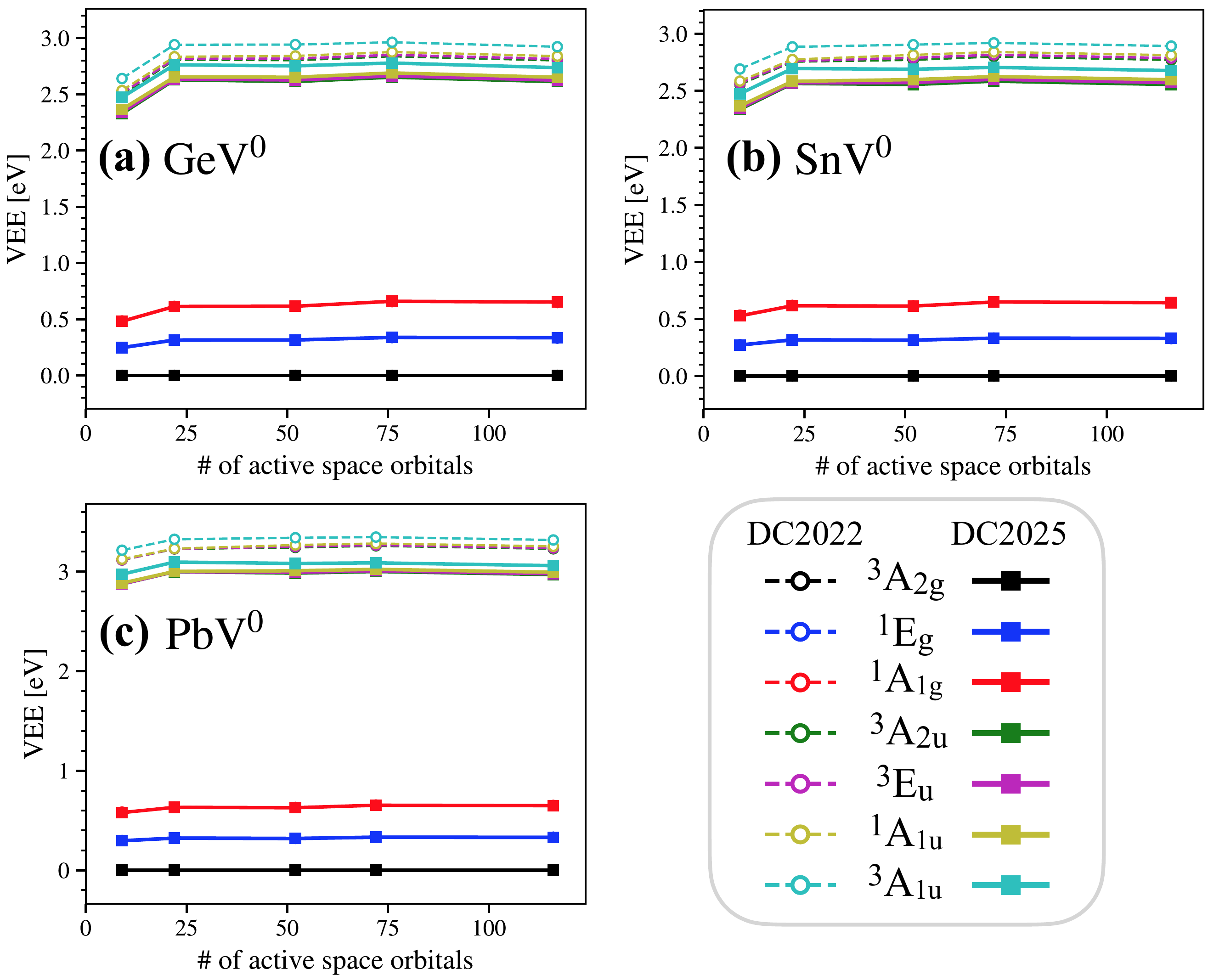}
\caption{\label{fig:supp_newDC_groupIV_511}
Comparison of computed many-body states using the DC2022 (dashed lines) and DC2025 (solid lines) double-counting terms, for a 511-atom supercell for the neutral group IV vacancies in diamond.
Vertical excitation energies (VEEs) are reported as a function of the number of orbitals in the active space.
}
\end{figure}

\begin{figure}
\includegraphics[width=0.7\linewidth]{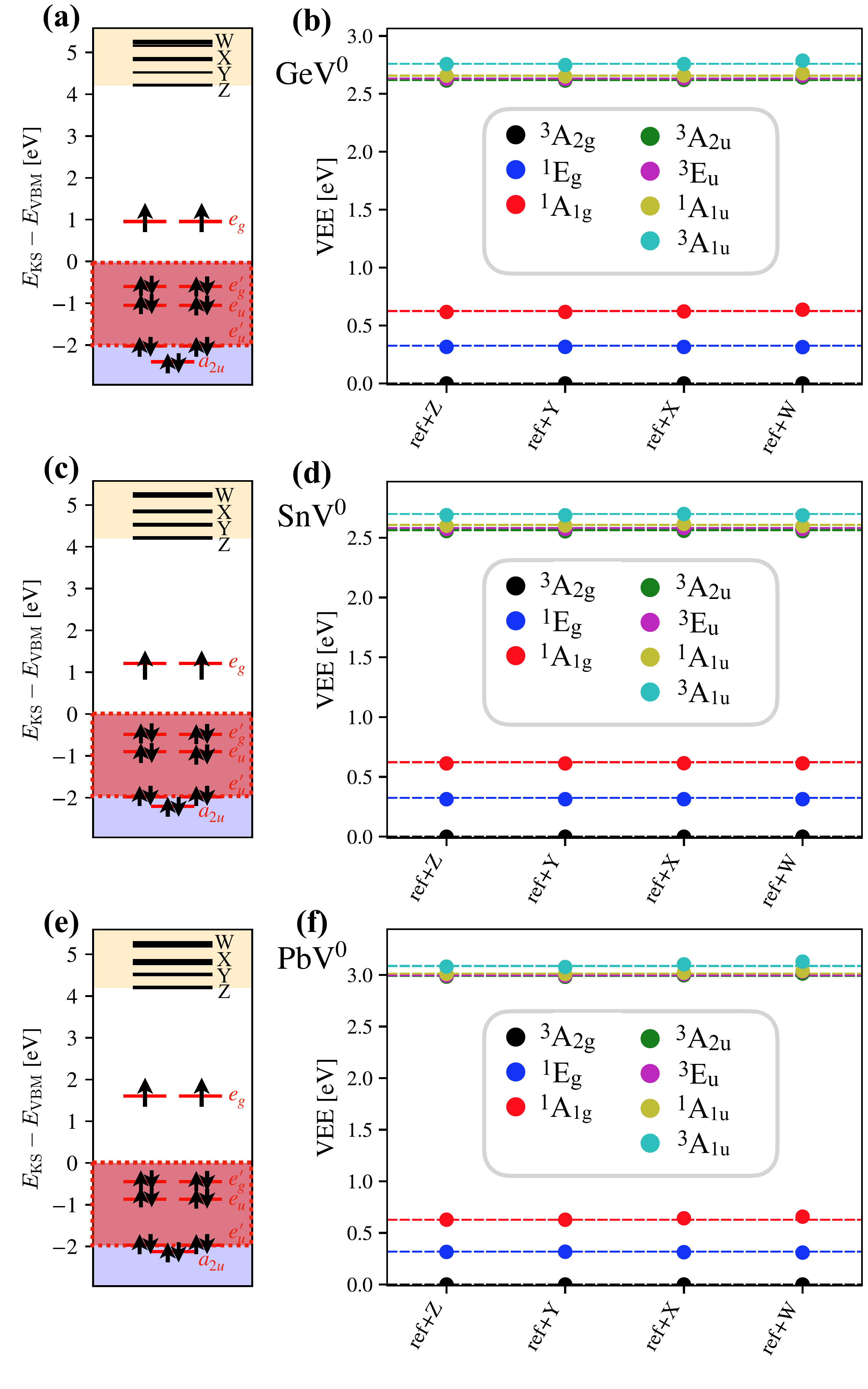}
\caption{\label{fig:supp_groupIV_unocc_effect}
Effect on computed vertical excitation energies (VEEs) of adding unoccupied orbitals into the active space, in 511-atom supercell calculations for the neutral group IV vacancies in diamond. (a)(c)(e) illustrate the chosen reference active space (``ref'') by red lines and red shades; it also shows groups of unoccupied orbitals added to the reference active space by letters. $E_\mathrm{KS}$ and $E_\mathrm{VBM}$ denote Kohn-Sham (KS) eigenvalues and the valence band maximum (VBM), respectively. (b)(d)(f): Comparison of VEEs computed with different active spaces, which are constructed by adding one group of unoccupied KS orbitals to a reference active space without unoccupied orbitals. The reference active space in all defects is the minimum model plus occupied orbitals with KS energies above VBM$-$2 eV. VEEs of the reference active space are marked with dashed horizontal lines. The MR-CISD impurity solver is used throughout.
}
\end{figure}

\begin{figure}
\includegraphics[width=0.9\linewidth]{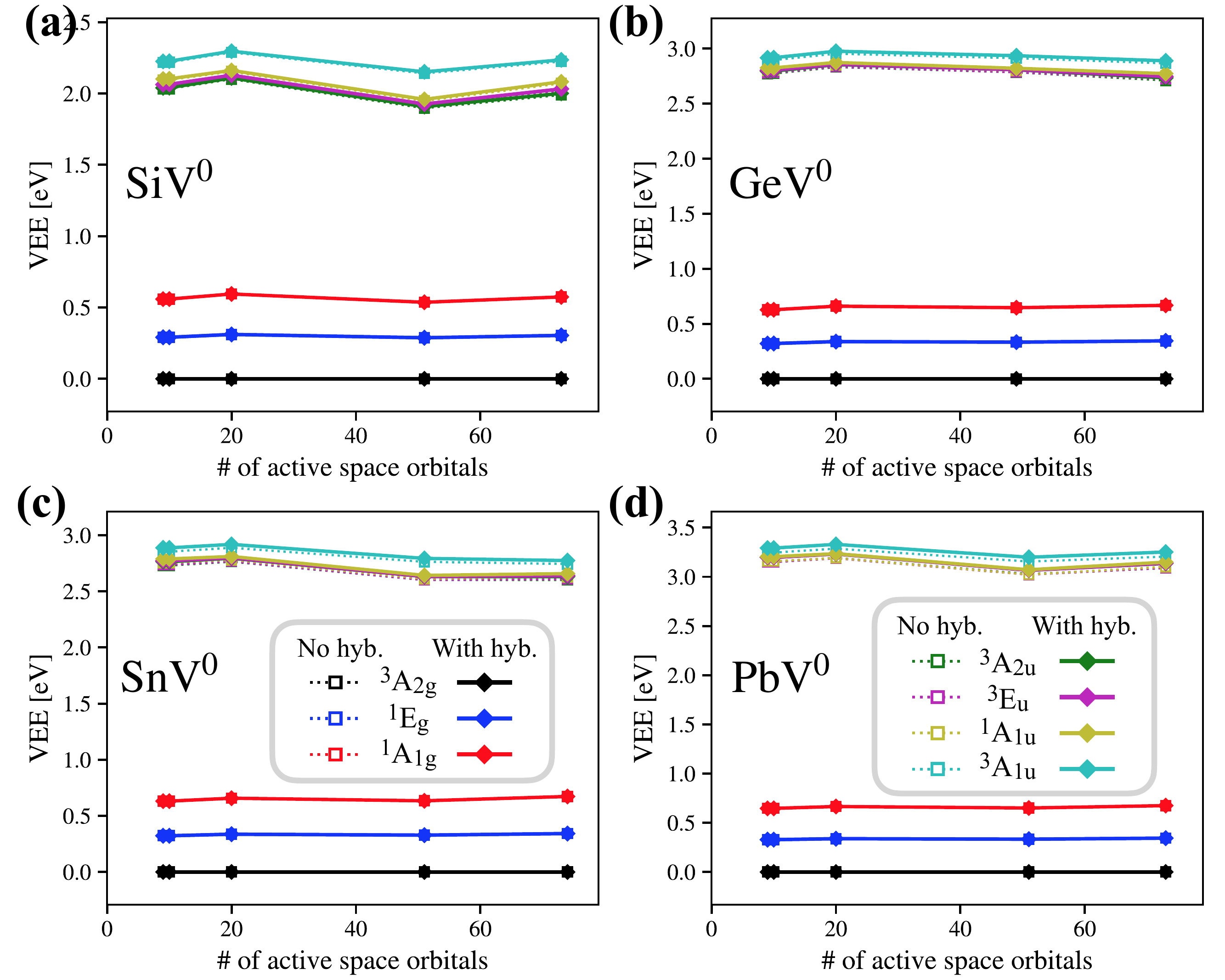}
\caption{\label{fig:supp_newDChyb_groupIV}Comparison of the vertical excitation energies without hybridization (dotted lines) and with hybridization (solid lines), as a function of the number of orbitals in the active space, for the NV$^-$ in diamond, computed with a 215-atom supercell. DFT calculations are carried out with the PBE functional and we used the refined double counting (DC2025). We used FCI and MR-CIS(D) as impurity solvers in the absence and presence of hybridization, respectively.
}
\end{figure}

Fig.~\ref{fig:supp_newDC_groupIV_511} shows the convergence of our results as a function of the size of the active space when comparing the double-counting schemes (DC2022 and DC2025, see main text), in a 511-atom supercell for $\mathrm{GeV^0}$, $\mathrm{SnV^0}$, and $\mathrm{PbV^0}$.
The results show a similar trend compared to those reported for $\mathrm{SiV^0}$ -- the highest excited states show a decrease of $\sim$0.2 eV in vertical excitation energies.
Fig.~\ref{fig:supp_groupIV_unocc_effect} shows the effect of unoccupied bands for $\mathrm{GeV^0}$, $\mathrm{SnV^0}$, and $\mathrm{PbV^0}$ in diamond in a 511-atom supercell.
Similar to the $\mathrm{SiV^0}$ defect, the effect of unoccupied bands is negligible.
Fig.~\ref{fig:supp_newDChyb_groupIV} shows the convergence of our results as a function of the size of the active space for all four neutral group IV vacancies with/without hybridization, in 215-atom supercells.
No splitting of states is observed for these defects, and the effect of hybridization is negligible throughout all active space choices.

\section{State splitting due to hybridization in NV$^-$@Diamond}

\begin{figure}
\includegraphics[width=0.8\linewidth]{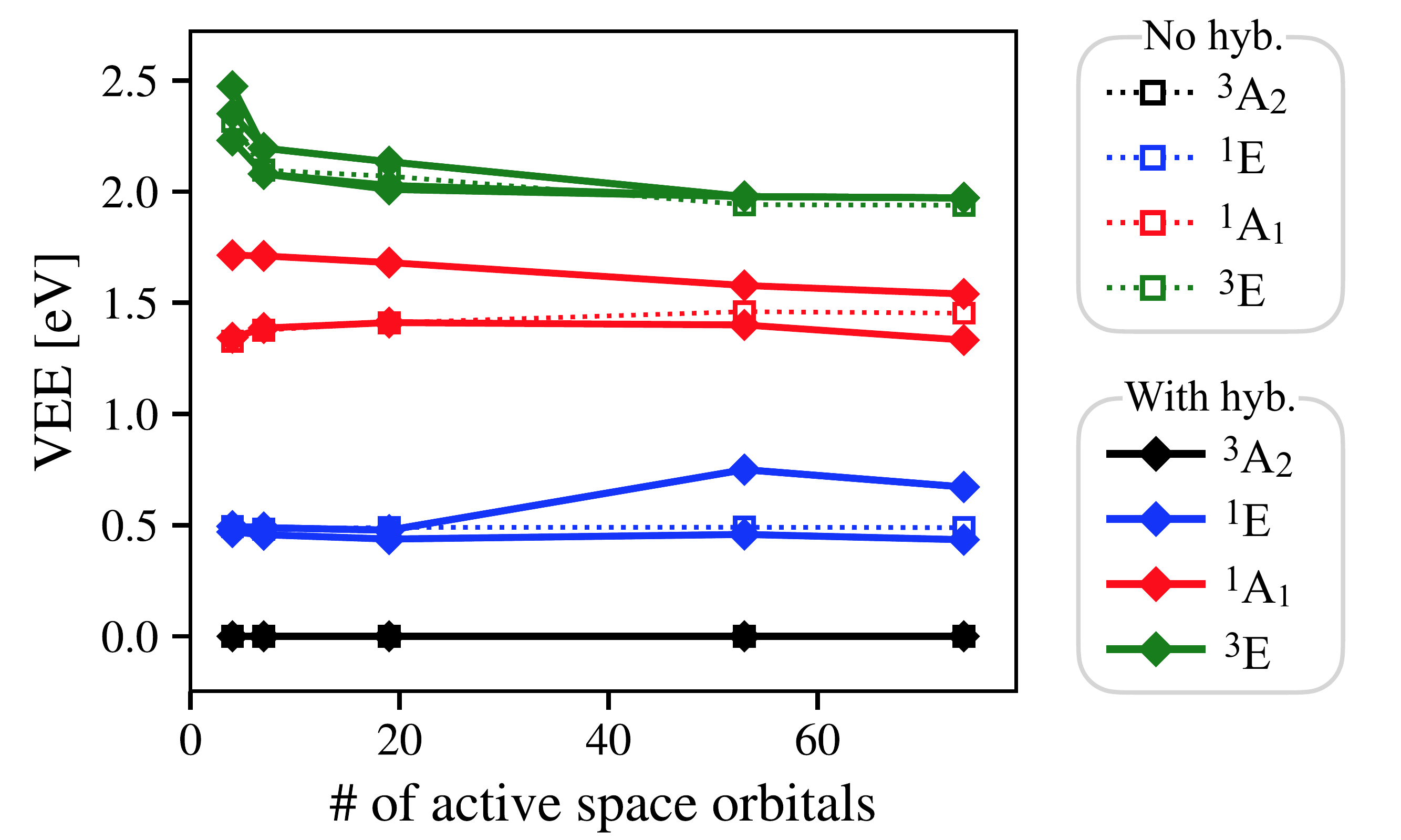}
\caption{\label{fig:supp_hyb_NV_split}Splitting of states due to hybridization in a 215-atom cell for the NV$^-$ center in diamond. Vertical excitation energies (VEEs) are plotted as a function of the size of the active space. Dotted lines show VEEs of states without hybridization, while solid lines show VEEs of all states that have more than 5\% component of the target symmetry state.}
\end{figure}

In Fig.~\ref{fig:supp_hyb_NV_split} we show the state splitting as a result of adding hybridization, for the NV$^-$ in diamond, computed with a 215-atom supercell. All states with more than 5\% component (in squared amplitude) of a given target symmetry state (represented by each color) are plotted. A split into 2--4 states is observed for some target symmetry state. In Section 4.4 of the main text, this state splitting is resolved by two approaches: choosing the lowest-energy state for each symmetry, or performing an ``ensemble average'' for all states for a given target symmetry state to get a unique VEE.

\section{Comment on excited state AFQMC calculations}

\CHANGES{
For excited-state calculations with AFQMC, the following complexities arise:
\begin{itemize}
    \item If the trial state of a given excited state has a small projection on an orthogonal ground or excited state of different energy, the AFQMC energy will slowly drift towards the energy of the lower-energy state during imaginary time propagation, even if the projection is minuscule (e.g. $10^{-10}$ value of the matrix element). AFQMC measurements must take place before the energy drifts too much, which requires a \textit{maximum} imaginary time cut-off.
    \item The CISD trial wave function is an eigenstate of the CISD approximate Hamiltonian, not of the many-body Hamiltonian used in AFQMC. Hence, the variational energy (the starting point of the AFQMC imaginary-time propagation) is different from the CISD energy obtained with the same wavefunction, possibly not as accurate. During imaginary time propagation, this variational energy equilibrates towards the true energy of the many-body state.
   AFQMC measurements can only take place after the energy is visually converged, requiring a \textit{minimum} imaginary time cutoff.
\end{itemize}

The issues outlined above imply that the imaginary time range for measurements must be carefully bounded on both sides, in our case by manually inspecting the total energy curve, to guarantee convergence while keeping the amplitude of any state orthogonal to the CISD trial wavefunction small. Due to the aforementioned energy drifting, the choice of the bounds introduces precision errors (estimated to be of the order of $10^{-2}$ eV) in addition to statistical errors.
}

\bibliography{supp}